\documentclass[11pt]{article}
\usepackage{amsmath}
\usepackage{amssymb}
\usepackage{graphicx}
\makeatletter
\usepackage{jheppub}
\renewcommand{\[}{\begin{equation}}
\renewcommand{\]}{\end{equation}}
\numberwithin{equation}{section}
\def\D{{\Delta}}

\def\CD{{\cal D}}

\def\CL{{\cal L}}
\def\CR{{\cal R}}

\def\CN{{\cal N}}

\def\CS{{\cal S}}

\makeatother
\begin{document}

\preprint{PUPT-2448}
\title{Supersymmetric R{\' e}nyi Entropy}
\author{Tatsuma Nishioka and Itamar Yaakov}
\affiliation{Department of Physics, Princeton University, \\ Princeton NJ 08544, USA}
\emailAdd{nishioka@princeton.edu}
\emailAdd{iyaakov@princeton.edu}
\abstract{We consider 3d $\mathcal{N}\ge 2$ superconformal field theories on a branched covering of a three-sphere. The  R{\' e}nyi entropy of a CFT is given by the partition function on this space, but conical singularities break the supersymmetry preserved in the bulk. We turn on a compensating $R$-symmetry gauge field and compute the partition function using localization. We define a supersymmetric observable, called the super R{\' e}nyi entropy, parametrized by a real number $q$. We show that the super R{\' e}nyi entropy is duality invariant and reduces to entanglement entropy in the $q\to 1$ limit.
We provide some examples.
}

\maketitle

\section{\label{sec:Introduction}Introduction}

Supersymmetric gauge theories provide a vast playground in which to
explore aspects of strongly coupled quantum field theory which are
not otherwise accessible. These models are generic enough to
incorporate the types of fields and interactions included in familiar
non-supersymmetric theories, and constrained enough to allow, for example,
exact evaluation of special observables even when perturbation theory
is inapplicable. Moreover, the ratio in this mixture can frequently
be tuned by changing the number of supercharges preserved by the model
and the observable.

The entanglement entropy of the vacuum is an example of an observable
which is available in any quantum field theory. To define this quantity,
one divides a spacial slice into a region $V$ and its complement
$-V$ (we use the notation of \cite{Casini:2009sr}). The Hilbert
space is likewise split 
\begin{equation}
\mathcal{H}=\mathcal{H}_{V}\otimes\mathcal{H}_{-V}\ .\label{eq:Hilbert_space_splitting}
\end{equation}
The reduced density matrix associated with $V$ of the vacuum state
$|0\rangle$ is defined by 
\[
\rho_{V}=\text{tr}_{-V}|0\rangle\langle0|\ ,
\]
where the trace is over a basis in $\mathcal{H}_{-V}$: the Hilbert
space of the states associated with field configurations supported on
$-V$. Normalization of the state $|0\rangle$ implies 
\[
\text{tr}_{V}\left(\rho_{V}\right)=1\ .
\]
The entanglement entropy is the von Neumann entropy of $\rho_{V}$
\[
S\left(V\right)=-\text{tr}_{V}\left(\rho_{V}\log\rho_{V}\right)\ .
\]
This provides a measure of the degree to which $|0\rangle$ departs
from a product state in the decomposition defined by \eqref{eq:Hilbert_space_splitting}.
$S\left(V\right)$ must be regularized and generically contains both
universal and non-universal parts. Universal components of $S\left(V\right)$ for a hyper-spherical
region have been shown to be related to $c$-functions for quantum
field theories in various dimensions \cite{Myers:2010tj}. A $c$-function
is an observable which is positive and monotonically decreasing along
RG flows, the prototypical example being the central charge of a 2d
conformal field theory when appropriately extended away from the fixed
points (see \cite{Casini:2009sr,Casini:2011kv} and references therein).
In 3d, $S\left(V\right)$ has only power law
divergences and the finite piece is universal. 

One can also consider the R{\' e}nyi entropies defined by 
\begin{equation}
S_{q}\left(V\right)=\frac{1}{1-q}\log\text{tr}_{V}\rho_{V}^{q}\ .\label{eq:Renyi_entropy}
\end{equation}
These contain additional information and can be used to further characterize
the ground state. The entanglement entropy can be recovered as 
\[
\lim_{q\rightarrow1}S_{q}\left(V\right)=S\left(V\right)\ .
\]
Indeed, this is often the easiest way to calculate $S\left(V\right)$.
This is because the calculation of $S_{q}$, for integer $q$, can
be reduced to the evaluation of a Euclidean partition function $Z_q$ using
the ``replica trick''
\[
S_{q}=\frac{1}{1-q}\log\left(\frac{Z_{q}}{\left(Z_{1}\right)^{q}}\right)\ ,
\]
where the functional path integral is performed
on a $q$-covering space branched along $V$. This reproduces both
the traces and the powers of the density matrix in \eqref{eq:Renyi_entropy}.
The $q$-covering space has curvature singularities along the boundary
$\partial V$. If the theory has conformal symmetry in $d$-dimensions,
the calculation can be mapped to a thermal partition function,
with temperature set by $q$, on either $\mathbb{H}^{d-1}\times\mathbb{R}$
or $dS_{d}$ \cite{Hung:2011nu}. The result is again equivalent
to a Euclidean partition function on either $\mathbb{H}^{d-1}\times S^{1}$,
where the circumference of the thermal circle is related to the curvature
of the hyperbolic space as $2\pi Rq$, or a branched covering of $S^{d}$.
The value of the entanglement entropy can be recovered by taking the
limit $q\rightarrow1$, or directly by evaluating the partition function
on either $S^{d}$ or the space $\mathbb{H}^{d-1}\times S^{1}$. The
calculation of $S_{q}$ has been carried out in two-dimensions \cite{Calabrese:2004eu,Casini:2005rm,Casini:2005zv,Azeyanagi:2007bj,Calabrese:2009ez,Calabrese:2010he,Headrick:2010zt,Headrick:2012fk,Lewkowycz:2012qr,Herzog:2013py,Hartman:2013mia,Faulkner:2013yia}
and for free fields in higher-dimensions \cite{Casini:2010kt,Klebanov:2011uf,Fursaev:2012mp}.
A calculation of the entanglement entropy for 3d conformal field theories
in the large-$N$ limit was presented in \cite{metlitski2009entanglement,Klebanov:2011td} and
a proposal for theories with holographic duals was put forth in \cite{Ryu:2006bv,Ryu:2006ef}.
In general, however, the calculation of either the entanglement or
R{\' e}nyi entropies for interacting theories, even ones with conformal
symmetry, is difficult.

Surprisingly, the Euclidean partition function on $S^{3}$ for a conformal
theory with $\mathcal{N}\ge2$ supersymmetry in the 3d sense (four supercharges)
can be recovered from a supersymmetric calculation. By this we mean
that there exists an action on $S^{3}$ for such a theory which preserves
a symmetry group that includes fermionic generators \cite{Festuccia:2011ws,Closset:2012ru}.
These can be used to localize the theory and reduce the calculation
of the partition function, indeed any supersymmetric observable, to
a finite matrix model \cite{Kapustin:2009kz,Jafferis:2010un,Hama:2010av}.
In practice, the finite part of the partition function is independent
of many  parameters in the effective action such as the D-terms. This
means that one can evaluate the partition function of a superconformal
field theory by embedding it in the flow from an appropriate UV action.
This allows one to compute the entanglement entropy even for strongly
coupled IR fixed points of 3d supersymmetric gauge theories as long
as one can identify a conserved $R$-symmetry current. This has been
used to test the conjecture that the entanglement entropy, or equivalently
the $S^{3}$ partition function, is a $c$-function \cite{Jafferis:2011zi,Klebanov:2011gs,Amariti:2011da,Morita:2011cs}.
This conjecture, now known as the $F$-theorem, is supported by the
$F$-maximization principle for $\CN=2$ supersymmetric field theories
\cite{Closset:2012vg}. A proof has been presented for general theories
with the use of the strong subadditivity property of entanglement
entropy \cite{Casini:2012ei}.

We will explore the possibility of recovering the R{\' e}nyi entropies
of a 3d $\mathcal{N}=2$ superconformal theory by calculating a similar
Euclidean partition function on the branched covering of $S^{3}$.\footnote{We would like to thank Rob Myers for suggesting this possibility.
} This space is homeomorphic to $S^3$ and also locally isometric to $S^{3}$
away from a certain co-dimension two singularity. The singularity
is supported on a great circle. It is conical in nature and represents
a deficit/surplus angle for $q<1$ and $q>1$ respectively. The main
challenge will be to regulate this singularity in a manner compatible
with supersymmetry. We will find that this requires turning on an
additional background value for the $R$-symmetry gauge field in the
supergravity multiplet. We call the branched sphere with this additional
background the singular space.

In section \ref{sec:Setup}, we show how to deal with the singularities
of the branched sphere in a supersymmetric way. 
We define a quantity which is related
to the R{\' e}nyi entropy of a 3d superconformal theory that we call
the super R{\' e}nyi entropy $S_{q}^{\text{susy}}$ by 
\[
S_{q}^{\text{susy}}=\frac{1}{1-q}\Re\left[ \log\left(\frac{Z_{\text{singular space}}\left(q\right)}{\left( Z_{S^3}\right)^{q}}\right) \right]\ .
\]
In section \ref{sec:Localization} we show
how to compute this quantity for a superconformal $\mathcal{N}=2$
theory using localization. We also consider more general situation where
the branched sphere is parametrized by two parameters $p$ and $q$
whose metric is given by \eqref{eq:two_parameter_metric}.

The localized partition function we obtain is summarized in the
following form: 
\begin{align}
Z(p,q)=\frac{1}{|W|}\int\prod_{i=1}^{\text{rank}\, G}d\sigma_{i}\, e^{\pi ik\,\text{Tr}(\sigma^{2})}\cdot\prod_{\alpha}\frac{1}{\Gamma_{h}\left(\alpha(\sigma)\right)}\cdot\prod_{I}\prod_{\rho\in\CR_{I}}\Gamma_{h}\left(\rho(\sigma)+i\omega\D_{I}\right)\ ,\label{eq:MMpq}
\end{align}
where 
\begin{align}
\Gamma_{h}\left(z\right)\equiv\Gamma_{h}\left(z;i\omega_{1},i\omega_{2}\right) \ ,
\end{align}
is the hyperbolic gamma function with 
\begin{align}
\omega_{1}=\sqrt{\frac{q}{p}} \ , \qquad \omega_{2}=\sqrt{\frac{p}{q}} \ , \qquad \omega=\frac{\omega_{1}+\omega_{2}}{2} \ , 
\end{align}
$k$ is the Chern-Simons level, $I$ labels the types of
chiral multiplets and $\rho$ is a weight in a representation $\CR_{I}$
of a gauge group $G$. $\D_{I}$ is the $R$-charge of the scalar
field in a chiral multiplet. The super R{\' e}nyi entropy is obtained by the relation
$Z_{\text{singular space}}(q)\equiv Z(1,q)$. The form of the partition
function \eqref{eq:MMpq} is equivalent to that on the squashed three-sphere
$S_{b}^{3}$ \cite{Imamura:2011wg,Hama:2011ea} with squashing parameter
$b=\sqrt{q/p}$\,. 

Section \ref{sec:Checks} contains several checks of the result, especially the duality invariance
and the interpretation of the partition function on the branched sphere
as insertions of defect operators on the round sphere. Several examples
are presented in section \ref{sec:Examples} to demonstrate how to
compute the super R{\' e}nyi entropy and show how it behaves as a function
of $q$.

Our conventions are summarized in appendix \ref{sec:Conventions}.
Appendices \ref{sec:The-branched-sphere} and \ref{sec:The-resolved-space}
contain the details of the branched and resolved three-spheres. The
definition of the hyperbolic gamma function and useful formulae are
collected in appendix \ref{sec:Special-functions}.

\section{\label{sec:Setup}Setup}

We are interested in defining a quantity which is as close as possible
to the usual R{\' e}nyi entropy and can be evaluated exactly using
localization. To this end we employ the simplification present for
conformal field theories which allows us to rewrite the R{\' e}nyi
entropy as a partition function over a curved manifold: the branched
three-sphere. We also restrict ourselves to studying superconformal
theories with $\mathcal{N}\ge2$ supersymmetry in the 3d sense. Such
theories can be consistently coupled to a fixed gravitational background
while preserving supersymmetry. This is accomplished by coupling
a special supermultiplet containing the energy-momentum tensor to
supergravity and taking an appropriate limit. The procedure is described
in detail in \cite{Closset:2012ru}.

\subsection{The branched sphere}

The original description of the branched sphere in \cite{Klebanov:2011uf}
is with coordinates $\theta,\tau,\phi$ with periodicities
$\theta\in[0, \pi/2]\ ,\tau\in[0,2\pi q)$, and $\phi\in[0,2\pi)$.
We will rescale $\tau$
by $1/q$. The new line element (with the rescaled variable still
called $\tau$) is 
\begin{align}
\begin{aligned}ds^{2} & =\ell^2\left( d\theta^{2}+q^{2}\sin^{2}\theta d\tau^{2}+\cos^{2}\theta d\phi^{2}\right)\ ,\\
\theta\in[0, & \pi/2]\ ,\qquad\tau\in[0,2\pi)\ ,\qquad\phi\in[0,2\pi)\ ,
\end{aligned}
\label{eq:one_parameter_metric}
\end{align}
and we consider the two parameter generalization 
\begin{align}
\begin{aligned}ds^{2} & =\ell^2\left( d\theta^{2}+q^{2}\sin^{2}\theta d\tau^{2}+p^{2}\cos^{2}\theta d\phi^{2}\right)\ ,\\
\theta\in[0, & \pi/2]\ ,\qquad\tau\in[0,2\pi)\ ,\qquad\phi\in[0,2\pi)\ .
\end{aligned}
\label{eq:two_parameter_metric}
\end{align}

Near $\theta=0$ the Ricci scalar on this space has a delta function
singularity as required by the Gauss-Bonnet theorem 
\begin{equation}
R\simeq 2\frac{q-1}{q}\frac{\delta\left(\theta\right)}{\theta}\ ,\label{eq:curvature_singularity}
\end{equation}
and a similar expression involving $p$ near $\theta=\pi/2$ (see \eqref{eq:RSdelta}). These
are co-dimension two conical singularities. The singularities occur
on two great circles with linking number 1. Other aspects of this
geometry are summarized in appendix \ref{sec:The-branched-sphere}.

\subsection{\label{sub:A-singular-supersymmetric}A singular supersymmetric background}
The gravitational background described above as the setting for the computation
of the R{\' e}nyi entropy can be embedded into a background supergravity
multiplet. It is a priori clear that on the bulk of the space the
configuration can be extended to the one that is supersymmetric by turning
on appropriate fields in the multiplet as in the usual $S^{3}$ computation.
The curvature singularities can then be viewed as gravitational defects.
These, too, can be made supersymmetric by including a compensating
gauge defect in the background gauge fields (specifically the $R$-symmetry
gauge field). One must still specify boundary conditions on the dynamical
fields of the theory as one approaches the singularity. 
From now on we will set $\ell=1$. Factors of $\ell$ can be easily restored,
however, the calculation of the super R{\' e}nyi entropy using localization
does not depend on it.

We wish to identify a supersymmetric action on the branched sphere
with metric \eqref{eq:two_parameter_metric} for a general $\mathcal{N}=2$
theory with a conserved $R$-symmetry. This was carried out for a
general background in \cite{Closset:2012ru} by coupling the theory
to 3d supergravity using the $\mathcal{R}$-multiplet of \cite{Komargodski:2010rb}.
The supergravity multiplet contains the bosonic fields 
\[
g_{\mu\nu}\ ,\qquad A_{\mu}\ ,\qquad V_{\mu}=-i{\varepsilon_{\mu}}^{\nu\rho}\partial_{\nu}C_{\rho}\ ,\qquad H=\frac{i}{2}\varepsilon^{\mu\nu\rho}\partial_{\mu}B_{\nu\rho}\ .
\]
These are defined up to gauge transformations which include diffeomorphisms
and background gauge transformations for the $R$-symmetry gauge field
$A_{\mu}$. We refer the reader to \cite{Closset:2012ru} for details.

To determine an appropriate supergravity background, we consider the
Killing spinor equation from \cite{Closset:2012ru} 
\begin{equation}
\left(\nabla_{\mu}-iA_{\mu}\right)\zeta=-\frac{1}{2}H\gamma_{\mu}\zeta-iV_{\mu}\zeta-\frac{1}{2}\varepsilon_{\mu\nu\rho}V^{\nu}\gamma^{\rho}\zeta\ ,\label{eq:Killing_Spinor_Equation}
\end{equation}
where the covariant derivative for a spinor is defined by 
\begin{align*}
\nabla_{\mu}\zeta & =\left(\partial_{\mu}+\frac{1}{8}{\omega_{\mu}}^{ij}\left[\sigma_{i},\sigma_{j}\right]\right)\zeta\ .
\end{align*}
A solution of this equation, with particular values for the supergravity
background fields $A_{\mu},V_{\mu},H$, represents a single complex
supersymmetry generated by the spinor $\zeta$ with $R$-charge
$+1$. A similar equation exists for spinors of $R$-charge $-1$
\begin{equation}
\left(\nabla_{\mu}+iA_{\mu}\right)\zeta=-\frac{1}{2}H\gamma_{\mu}\zeta+iV_{\mu}\zeta+\frac{1}{2}\varepsilon_{\mu\nu\rho}V^{\nu}\gamma^{\rho}\zeta\ .\label{eq:Killing_Spinor_Equation_Negative_R_Charge}
\end{equation}
On $S^{3}$, one can preserve two spinors of each $R$-charge by turning
on, in addition to the usual metric for the round sphere, a constant
imaginary value for the field $H=\pm i$.%
\footnote{Since $A_{\mu}=V_{\mu}=0$ in this setting, the equations \eqref{eq:Killing_Spinor_Equation}
and \eqref{eq:Killing_Spinor_Equation_Negative_R_Charge} coincide.
The spinors used to generate supersymmetries with positive and negative
$R$-charges have the same spatial profile.%
}

The usual spinor used in localization on $S^{3}$, $\varepsilon_{L}$,
satisfies (in the left-invariant basis in appendix \ref{sec:The-branched-sphere})
\[
\nabla_{\mu}^{S^{3}}\varepsilon_{L}=\left(\partial_{\mu}+\frac{1}{8}{\omega_{S^{3}\mu}}^{ij}\left[\sigma_{i},\sigma_{j}\right]\right)\varepsilon_{L}=\frac{i}{2}{\left(e_{S^{3}}\right)_{\mu}}^{i}\sigma_{i}\varepsilon_{L}\ ,
\]
because in the left-invariant basis on $S^{3}$ the spin connection
is related to the vielbein as 
\[
\omega^{ij}|_{q=p=1}=\varepsilon^{ijk}e_{k}|_{q=p=1}\ ,
\]
and $\varepsilon_{L}$ is (any) constant spinor in this basis \cite{Kapustin:2009kz}.
There are two more spinors (which are constant in the right-invariant
basis) that satisfy 
\[
\nabla_{\mu}^{S^{3}}\varepsilon_{R}=\left(\partial_{\mu}+\frac{1}{8}{\omega_{S^{3}\mu}}^{ij}\left[\sigma_{i},\sigma_{j}\right]\right)\varepsilon_{R}=-\frac{i}{2}{\left(e_{S^{3}}\right)_{\mu}}^{i}\sigma_{i}\varepsilon_{R}\ .
\]
On the branched sphere, the spin connection satisfies
\[
\omega^{ij}=\varepsilon^{ijk}e_{k}-\tilde{\omega}^{ij}\ ,
\]
where 
\[
\tilde{\omega}^{ij}=\varepsilon^{ij3}\left(\left(q-1\right)d\tau+\left(p-1\right)d\phi\right)\ .
\]

\subsubsection{Bulk}

For the branched sphere with metric \eqref{eq:two_parameter_metric},
the spinor equation can be satisfied in the bulk by taking 
\[
H=-i\ ,\qquad A=0\ ,\qquad V=0\ ,
\]
and the spinors 
\[
\varepsilon_{\text{branched}}^{\pm}=\left(\begin{array}{c}
c_{+}e^{i\frac{q-1}{2}\tau+i\frac{p-1}{2}\phi}\\
c_{-}e^{-i\frac{q-1}{2}\tau-i\frac{p-1}{2}\phi}
\end{array}\right)\ .
\]
This follows from the fact that this space is (locally) isometric
to $S^{3}$ and these are nothing but the constant spinors of $\varepsilon_{L}$
after a coordinate transformation.

However, the conical singularities \eqref{eq:curvature_singularity} make this
solution suspect near $\theta=0,\pi/2$. Note that the equation (5.8)
of \cite{Closset:2012ru} implies the integrability condition (at
constant $H$ and $V=0$) 
\begin{equation}
\left[\frac{i}{2}\left(2R_{\mu\nu}-Rg_{\mu\nu}\right)\gamma^{\nu}-2i\varepsilon_{\mu\nu\rho}\nabla^{\nu}A^{\rho}+iH^{2}\gamma_{\mu}\right]\zeta=0\ ,\label{eq:Integrability_condition}
\end{equation}
which would include delta function contributions from \eqref{eq:curvature_singularity}.

\subsubsection{Singularities}

A somewhat better solution is to take 
\begin{equation}
H=-i\ ,\qquad A=\frac{q-1}{2}d\tau+\frac{p-1}{2}d\phi\ ,\qquad V=0\ ,\label{eq:SUGRA_background}
\end{equation}
so that the field strength 
\begin{equation}
F=\frac{q-1}{2}\delta\left(\theta\right)d\theta\wedge d\tau+\frac{p-1}{2}\delta\left(\frac{\pi}{2}-\theta\right)d\theta\wedge d\phi\ ,\label{eq:Singular_Field_Strength}
\end{equation}
compensates for the curvature singularity. Near the singularities
we have the identities that satisfy the integrability condition \eqref{eq:Integrability_condition}
\begin{align}
\begin{aligned}\left[\frac{1-p}{2}\delta\left(\frac{\pi}{2}-\theta\right)\sigma_{3}+F_{\phi\theta}\right]\zeta & =0\ ,\\
\left[\frac{1-q}{2}\delta\left(\theta\right)\sigma_{3}-F_{\theta\tau}\right]\zeta & =0\ ,
\end{aligned}
\label{IntCond}
\end{align}
and the additional background gauge field cancels the additional term in the spin connection
\[
\frac{1}{8}{\tilde{\omega}_{\mu}}^{ij}\left[\sigma_{i},\sigma_{j}\right]=\frac{i}{2}\left({\delta_{\mu}}^{\tau}\left(q-1\right)+{\delta_{\mu}}^{\phi}\left(p-1\right)\right)\sigma_{3}\ ,
\]
in equations \eqref{eq:Killing_Spinor_Equation} and \eqref{eq:Killing_Spinor_Equation_Negative_R_Charge} 
provided we choose the $R$-charge correctly. This amounts to performing
a singular background gauge transformation for the $R$-symmetry.

This setup preserves two constant spinors, in the basis defined by
\eqref{eq:vielbein}, which satisfy the equations

\[
\sigma_{3}\zeta_{\pm}=\pm\zeta_{\pm}\ ,
\]

\[
\left(\nabla_{\mu}\pm iA_{\mu}\right)\zeta_{\pm}=-\frac{1}{2}H\gamma_{\mu}\zeta_{\pm} \ ,
\]
so that the associated $R$-charges are
\[
r\left(\zeta_{\pm}\right)=\mp1 \ .
\]
There is actually a four complex parameter family of spinors preserved
by this background. The basis spinors are given by 
\begin{align}
\begin{aligned}\zeta_{+} & =\left(\begin{array}{c}
c_{+}^{1}\\
c_{+}^{2}e^{i\left(1-q\right)\tau+i\left(1-p\right)\phi}
\end{array}\right),\qquad r=-1\ ,\\
\zeta_{-} & =\left(\begin{array}{c}
c_{-}^{1}e^{i\left(q-1\right)\tau+i\left(p-1\right)\phi}\\
c_{-}^{2}
\end{array}\right),\qquad r=1\ .
\end{aligned}
\end{align}
Note that the spinors parametrized by $c_{+}^{2}$ and $c_{-}^{1}$
still satisfy \eqref{eq:Integrability_condition} (taking into account
the $R$-charge) only up to delta function contributions.

The fact that this background preserves four independent supercharges
is in conflict with expectations. Specifically, a smooth background
preserving four supercharges always has constant $H$ and the field
strength for the connection $A_{\mu}-V_{\mu}$ is flat \cite{Closset:2012ru}.
The second condition is not satisfied for \eqref{eq:SUGRA_background},
which implies that we may have used the Killing spinor equation \eqref{eq:Killing_Spinor_Equation}
outside its range of validity. For one, we have ignored possible boundary
terms used in its derivation.%
\footnote{We would like to thank Thomas Dumitrescu for stressing this to us.%
} We could attempt to fix this by excising the loops at $\theta=0,\pi/2$
and extending the formalism in \cite{Closset:2012ru} to include spaces
with a boundary. We will instead take a simpler approach which allows
us to continue working with a compact space.

\subsection{\label{sub:The-resolved-space}The resolved space}

We can avoid the pitfalls of the singular background by smoothing
out both the curvature of \eqref{eq:curvature_singularity} and the
field strength of \eqref{eq:Singular_Field_Strength}. We do this
by introducing a one parameter family of smooth backgrounds (the resolved
space) which solve the Killing spinor equation \eqref{eq:Killing_Spinor_Equation}
and converge to \eqref{eq:two_parameter_metric} and \eqref{eq:SUGRA_background}.
This can be achieved by deforming the metric and vielbein as follows

\begin{equation}
ds^{2}=\frac{1}{f_{\epsilon}\left(\theta\right)}d\theta^{2}+q^{2}\sin^{2}\theta d\tau^{2}+p^{2}\cos^{2}\theta d\phi^{2}\ ,\label{eq:smooth_two_parameter_metric}
\end{equation}
and introducing the following smooth background fields 
\begin{equation}
H=-i\sqrt{f_{\epsilon}\left(\theta\right)}\ ,\qquad A=\frac{q\sqrt{f_{\epsilon}\left(\theta\right)}-1}{2}d\tau+\frac{p\sqrt{f_{\epsilon}\left(\theta\right)}-1}{2}d\phi\ ,\qquad V=0\ ,\label{eq:smooth_SUGRA_background}
\end{equation}
where $f_{\epsilon}\left(\theta\right)$ is a smooth function satisfying
(for a small $\epsilon>0$) 
\[
f_{\epsilon}\left(\theta\right)=\begin{cases}
\frac{1}{q^{2}}\ , & \theta\rightarrow0 \ ,\\
\frac{1}{p^{2}}\ , & \theta\rightarrow\frac{\pi}{2} \ ,\\
1\ , & \epsilon<\theta<\frac{\pi}{2}-\epsilon \ .
\end{cases}
\]
As $\epsilon\rightarrow0$ this approaches the required background.
We will henceforth write simply $f\left(\theta\right)$.

The resolved space still preserves two of the spinors of the singular
space 
\begin{align}
\zeta_{+} & =\left(\begin{array}{c}
c_{+}^{1}\\
0
\end{array}\right),\qquad r=-1\ ,\label{eq:Smooth_supersymmetry_1}\\
\zeta_{-} & =\left(\begin{array}{c}
0\\
c_{-}^{2}
\end{array}\right),\qquad r=1\ ,\label{eq:Smooth_supersymmetry_2}
\end{align}
which will be enough to perform the localization on this space in the next section. The boundary conditions for the modes of
fields defined on this background are the usual ones which produce
non-singular and normalizable field configurations. Since we are interested
in the original partition function defining the super R{\' e}nyi
entropy, we will take $\epsilon\rightarrow0$ at the end of the localization
procedure. 

Although the partition function on the resolved space could, in principle,
depends on the detailed form of the deformation function $f\left(\theta\right)$,
the limit will not. One may still wonder whether
there are different supersymmetric backgrounds which could have been
used to define the resolved space. For instance, one may consider
the more general configuration 
\begin{align}
\begin{aligned}
H_{h} & =  -ih\left(\theta \right)\ ,\\
A_{h} & =\frac{1}{4}\left(-2+q\left(-1+3\cos\left(2\theta\right)\right)\sqrt{f_{\epsilon}\left(\theta\right)}+6q\,h\left(\theta \right)\sin^{2}\theta\right)d\tau\\
	& \qquad +\frac{1}{4}\left(-2-p\left(1+3\cos\left(2\theta\right)\right)\sqrt{f_{\epsilon}\left(\theta\right)} +6p\,h\left(\theta\right)\cos^{2}\theta\right)d\phi\ ,\label{eq:smooth_SUGRA_background-1}\\
V_{h} & =  q\left(h\left(\theta\right)-\sqrt{f_{\epsilon}\left(\theta\right)}\right)\sin^{2}\theta d\tau+p\left(h\left(\theta\right)-\sqrt{f_{\epsilon}\left(\theta\right)}\right)\cos^{2}\theta d\phi \ ,
\end{aligned}
\end{align}
which represents an infinite class of smooth backgrounds, parametrized
by the function $h\left(x\right)$, with which one can take the limit
to the branched sphere. The spinors \eqref{eq:Smooth_supersymmetry_1}
and \eqref{eq:Smooth_supersymmetry_2} are preserved for any $h\left(\theta\right)$
and this is the most general configuration. However, as noted
in \cite{Closset:2012ru}, the scalar $H$ couples to an operator
which becomes redundant in a conformal theory and hence decouples
in the IR. The fields $A$ and $V$ still couple to the distinguished
$R$-symmetry current $j_{\mu}^{\left(R\right)}$ of a superconformal theory which sits in the same
multiplet as the energy-momentum tensor. The linearized coupling is
of the form 
\[
j_{\mu}^{\left(R\right)}\left(A^{\mu}-\frac{3}{2}V^{\mu}\right)\ ,
\]
but we can see that
\[
\frac{\delta}{\delta h\left(\theta\right)}\left(A_{h}^{\mu}-\frac{3}{2}V_{h}^{\mu}\right)=0 \ ,
\]
so a small change in $h\left(\theta\right)$ does not affect the result.
There may be more general configurations which would yield different
partition functions and hence a different type of supersymmetric observable
associated with the R{\' e}nyi entropy. We will not pursue this possibility.

One could try to form this type of a resolved space without the use
of the background $R$-symmetry gauge field. The Killing spinor equations then
imply that $f\left(\theta\right)\equiv1$ and hence the deformation
does not remove the singularities. We take this to mean that supersymmetry
on the branched sphere requires the introduction of the background
$R$-symmetry gauge field or the equivalent boundary conditions. Hence,
the partition function yielding the usual R{\' e}nyi entropy, with
boundary conditions of the type described in \cite{Klebanov:2011uf},
is not a supersymmetric observable.

\section{\label{sec:Localization}Localization}

The path integral expression which calculates the partition function
on the space defined by \eqref{eq:smooth_two_parameter_metric} and
\eqref{eq:smooth_SUGRA_background} is invariant under the fermionic
symmetries generated by \eqref{eq:Smooth_supersymmetry_1} and \eqref{eq:Smooth_supersymmetry_2}.
Let us call $Q$ the supercharge generated by $\zeta$. A standard argument
shows that the partition function can be expressed as a sum of contributions
from fixed points of $Q$ along with a determinant (or superdeterminant)
representing the equivariant Euler class of the normal bundle \cite{Witten:1999ds,Pestun:2007rz}.
One must also evaluate the classical action on this space. Details
are available in \cite{Marino:2011nm,Kapustin:2009kz}. We now carry
out this localization calculation by finding the fixed points of the
$Q$ action, the fluctuation determinants and the classical contributions.

\subsection{The localizing term}

An efficient way of finding the space of fixed points is to add a
$Q$-exact term $\{ Q, V\} $ to the action $S$ of the $\mathcal{N}=2$ theory under
consideration.
The same argument as above shows that the result is
independent of this term. 
The deformed partition function 
\begin{align}
Z(t) = \int \CD \phi \, e^{-S - t\{ Q, V\} } \ ,
\end{align}
does not depend on the parameter $t$. 
We will choose a positive semi-definite term with large $t$. The semi-classical approximation
around the zero locus is then exact.

We follow the supersymmetry transformation rules in \cite{Closset:2012ru}
to obtain an appropriate localizing term for vector and chiral
multiplets. A general superfield $\CS$ whose bottom component is
a complex scalar $C$ of $R$-charge $r$ and central charge $z$
has $16+16$ bosonic and fermionic components 
\begin{align}
\begin{aligned}\CS & =(C,\chi_\alpha, \tilde\chi_\alpha, M, \tilde M, a_\mu, \sigma, \lambda_\alpha, \tilde \lambda_\alpha, D)\ .
\end{aligned}
\end{align}
The supersymmetry transformation rules with respect to the Killing
spinor $\zeta$ of $R$-charge $+1$ are given by 
\begin{align}
\begin{aligned}\delta_{\zeta} C & =i\zeta\chi\ ,\\
\delta_{\zeta}\chi & =\zeta M\ ,\\
\delta_{\zeta}\tilde{\chi} & =-\zeta\left(\sigma-(z-rH)C\right)-\gamma^{\mu}\zeta(D_{\mu}C-ia_{\mu})\ ,\\
\delta_{\zeta}M & =0\ ,\\
\delta_{\zeta}\tilde{M} & =2\zeta\lambda-2i(z-(r+2)H)\zeta\tilde{\chi}-2iD_{\mu}(\zeta\gamma^{\mu}\tilde{\chi})\ ,\\
\delta_{\zeta}a_{\mu} & =-i\zeta\gamma_{\mu}\tilde{\lambda}+D_{\mu}(\zeta\chi)\ ,\\
\delta_{\zeta}\sigma & =-\zeta\tilde{\lambda}+i(z-rH)\zeta\chi\ ,\\
\delta_{\zeta}\lambda & =i\zeta(D+\sigma H)-i\varepsilon^{\mu\nu\rho}\gamma_{\rho}\zeta D_{\mu}a_{\nu}-\gamma^{\mu}\zeta\left((z-rH)a_{\mu}+iD_{\mu}\sigma-V_{\mu}\sigma\right)\ ,\\
\delta_{\zeta}\tilde{\lambda} & =0\ ,\\
\delta_{\zeta}D & =D_{\mu}(\zeta\gamma^{\mu}\tilde{\lambda})-iV_{\mu}\zeta\gamma^{\mu}\tilde{\lambda}-H\zeta\tilde{\lambda}+(z-rH)\left(\zeta\tilde{\lambda}-iH\zeta\chi\right)+\frac{ir}{4}(R-2V^{\mu}V_{\mu}-6H^{2})\zeta\chi\ ,
\end{aligned}
\label{eq:Susy_Rule_Zeta}
\end{align}
and for the Killing spinor $\tilde{\zeta}$ of $R$-charge $-1$ 
\begin{align}
\begin{aligned}\delta_{\tilde\zeta} C & =i\tilde{\zeta}\tilde{\chi}\ ,\\
\delta_{\tilde{\zeta}}\chi & =-\tilde{\zeta}\left(\sigma+(z-rH)C\right)-\gamma^{\mu}\tilde{\zeta}(D_{\mu}C+ia_{\mu})\ ,\\
\delta_{\tilde{\zeta}}\tilde{\chi} & =\tilde{\zeta}\tilde{M}\ ,\\
\delta_{\tilde{\zeta}}M & =-2\tilde{\zeta}\tilde{\lambda}+2i(z-(r-2)H)\tilde{\zeta}\chi-2iD_{\mu}(\tilde{\zeta}\gamma^{\mu}\chi)\ ,\\
\delta_{\tilde{\zeta}}\tilde{M} & =0\ ,\\
\delta_{\tilde{\zeta}}a_{\mu} & =-i\tilde{\zeta}\gamma_{\mu}\lambda-D_{\mu}(\tilde{\zeta}\tilde{\chi})\ ,\\
\delta_{\tilde{\zeta}}\sigma & =\tilde{\zeta}\lambda-i(z-rH)\tilde{\zeta}\tilde{\chi}\ ,\\
\delta_{\tilde{\zeta}}\lambda & =0\ ,\\
\delta_{\tilde{\zeta}}\tilde{\lambda} & =-i\tilde{\zeta}(D+\sigma H)-i\varepsilon^{\mu\nu\rho}\gamma_{\rho}\tilde{\zeta}D_{\mu}a_{\nu}+\gamma^{\mu}\tilde{\zeta}\left((z-rH)a_{\mu}+iD_{\mu}\sigma+V_{\mu}\sigma\right)\ ,\\
\delta_{\tilde{\zeta}}D & =-D_{\mu}(\tilde{\zeta}\gamma^{\mu}\lambda)-iV_{\mu}\tilde{\zeta}\gamma^{\mu}\lambda+H\tilde{\zeta}\lambda+(z-rH)\left(\tilde{\zeta}\lambda+iH\tilde{\zeta}\tilde{\chi}\right)-\frac{ir}{4}(R-2V^{\mu}V_{\mu}-6H^{2})\tilde{\zeta}\tilde{\chi}\ ,
\end{aligned}
\label{eq:Susy_Rule_Zeta_Bar}
\end{align}
where the covariant derivative is defined by 
\begin{align}
D_{\mu}=\nabla_{\mu}-ir\left(A_{\mu}-\frac{1}{2}V_{\mu}\right)-iz\,C_{\mu}\ .
\end{align}

A supersymmetric action can be obtained from the D-term (plus terms coupled to the background supergravity fields) of a general
superfield $\CS$ of $r=z=0$
\begin{align}
\CL_{D}=-\frac{1}{2}\left(D-\sigma H-a_{\mu}V^{\mu}\right)\ ,
\end{align}
whose supersymmetry transformation is a total derivative. One can
show that this action is $Q$-exact when $V_{\mu}=0$. More
explicitly, one can derive by using the transformation
rules (\ref{eq:Susy_Rule_Zeta}) and (\ref{eq:Susy_Rule_Zeta_Bar})\footnote{
Spinors are commuting variables and the supersymmetry transformation is anti-commuting
in \cite{Closset:2012ru}.
In this convention, two spinors $\xi$ and $\eta$ satisfy $\xi \eta = - \eta \xi$, $\xi \gamma_\mu \eta = \eta \gamma_\mu \xi$ and so on.
We would like to thank Guido Festuccia for clarifying this point for us.
}
\begin{align}
\begin{aligned}\delta_{\zeta}\delta_{\tilde{\zeta}}(i\sigma+2iHC) & =\zeta\tilde{\zeta}(D-\sigma H)-2iHK^{\mu}a_{\mu}-\varepsilon^{\mu\nu\rho}a_{\nu}\nabla_{\mu}K_{\rho} + i V^\mu K_\mu \sigma\\
 & \qquad+\nabla_{\mu}\left(\varepsilon^{\mu\nu\rho}a_{\nu}K_{\rho}+K^{\mu}\sigma + 2H C\right)\ ,
\end{aligned}
\label{eq:Q_exact_D_term}
\end{align}
where we used the notation 
\[
K^{\mu}\equiv\zeta\gamma^{\mu}\tilde{\zeta}\ ,
\]
that satisfies 
\[
\nabla^{\mu}K_{\mu}=0\ .
\]
The total derivative term in the second line of \eqref{eq:Q_exact_D_term}
vanishes inside the spacetime integral. One can show that the second and third terms in
the first line vanish
due to the following identity derived from the Killing spinor equations
\eqref{eq:Killing_Spinor_Equation} and \eqref{eq:Killing_Spinor_Equation_Negative_R_Charge}
\begin{align}
\nabla_{\mu}K_{\rho}=iH\varepsilon_{\mu\rho\kappa}K^{\kappa}+\varepsilon_{\mu\rho\kappa}V^{\kappa}\zeta\tilde{\zeta}\ .
\end{align}
Thus we have shown that the D-term of a general superfield of $r=z=0$
is always $Q$-exact as long as the background gauge field $V_{\mu}$
vanishes%
\footnote{The overall normalization is immaterial as long as the function $\zeta\tilde{\zeta}\ne0$.
In our case $\zeta\tilde{\zeta}\ne1$.%
} 
\begin{align}
\zeta\tilde{\zeta}\,\CL_{D}|_{r=z=0}=\delta_{\zeta}\delta_{\tilde{\zeta}}\left(-\frac{i}{2}\sigma-iHC\right)\ .\label{eq:D_Qext}
\end{align}
The relation \eqref{eq:D_Qext} ensures that the Yang-Mills and matter
actions given as D-terms
are $Q$-exact and can be used to localize the partition function
(see \cite{Closset:2012ru} for the vector and chiral multiplets and their transformation rules).
This also guarantees that the resulting partition function is independent of $\ell$, as a
change in $\ell$ can be absorbed, by rescaling the dynamical fields,
into the overall normalization of one of the $Q$-exact terms.

\subsection{\label{sub:Zero-modes}Zero modes}

The Yang-Mills term is $Q$-exact and can be used to localize the
gauge sector 
\begin{align}
\begin{aligned}\CL_{\text{YM}} & =\text{Tr}\Bigg[\frac{1}{4}f_{\mu\nu}f^{\mu\nu}+\frac{1}{2}D_{\mu}\sigma D^{\mu}\sigma-i\tilde{\lambda}\gamma^{\mu}\left(D_{\mu}+\frac{i}{2}V_{\mu}\right)\lambda\\
 & \quad+\frac{i}{2}\sigma\varepsilon^{\mu\nu\rho}V_{\mu}f_{\nu\rho}-\frac{1}{2}V_{\mu}V^{\mu}\sigma^{2}-\frac{1}{2}(D+\sigma H)^{2}-i\tilde{\lambda}[\sigma,\lambda]+\frac{i}{2}H\tilde{\lambda}\lambda\Bigg]\ ,
\end{aligned}
\label{eq:YMaction}
\end{align}
where 
\begin{align}
\begin{aligned}f_{\mu\nu} & =\partial_{\mu}a_{\nu}-\partial_{\nu}a_{\mu}-i[a_{\mu},a_{\nu}]\ ,\\
D_{\mu}\sigma & =\partial_{\mu}\sigma-i[a_{\mu},\sigma]\ ,\\
D_{\mu}\lambda & =(\nabla_{\mu}+iA_{\mu})\lambda-i[a_{\mu},\lambda]\ ,
\end{aligned}
\end{align}
and $a_{\mu}$ is a connection on a (trivial) principle bundle for a Lie group
$G$ with Lie algebra $\mathfrak{g}$ and all other fields are valued
in the adjoint bundle. We can fix the gauge freedom by adding the
gauge fixing term%
\footnote{We actually need to deal with a combined supersymmetry and BRST complex.
This extension is explained in \cite{Kapustin:2009kz}.%
} 
\begin{align}
\CL_{\text{g.f.}}=\bar{c}\,\nabla_{\mu}D^{\mu}c+b\,\nabla^{\mu}a_{\mu}\ ,\label{eq:Gauge_fixing_term}
\end{align}
which imposes the covariant gauge $\nabla^{\mu}a_{\mu}=0$. For $V_{\mu}=0$,
the bosonic part of the above action is a sum of positive semi-definite
terms. It is easy to see that it can vanish only when 
\begin{align}
\begin{aligned}f_{\mu\nu}=0\ ,\qquad D_{\mu}\sigma=0\ ,\qquad D=-H\sigma\ .\end{aligned}
\end{align}
The first condition leads to a flat connection which is equivalent
to $a_{\mu}=0$ on a smooth simply connected compact space. It follows
that the second condition gives the constant configuration of the
adjoint scalar 
\begin{align}
\sigma=\sigma_{0}\ ,
\end{align}
and the third yields 
\begin{align}
D=-H\sigma_{0}\ .
\end{align}
We refer to the Lie algebra valued variable $\sigma_{0}$ as a zero
mode.

The $Q$-exact term used to localize the matter fields of $R$-charge
$\D$ is 
\begin{align}
\begin{aligned}\CL_{\text{matter}} & ={\mathcal{D}}^{\mu}\tilde{\phi}{\mathcal{D}}_{\mu}\phi-i\tilde{\psi}\gamma^{\mu}{\mathcal{D}}_{\mu}\psi-\tilde{F}F\\
 & \quad+\left[D+\left(\sigma+\left(\D-\frac{1}{2}\right)H\right)^{2}-\frac{\D}{4}R+\frac{1}{2}\left(\D-\frac{1}{2}\right)(V^{\mu}V_{\mu}+H^{2})\right]\tilde{\phi}\phi\\
 & \qquad-\left(\sigma+\left(\D-\frac{1}{2}\right)H\right)i\tilde{\psi}\psi+\sqrt{2}i(\tilde{\phi}\lambda\psi+\phi\tilde{\lambda}\tilde{\psi})\ ,
\end{aligned}
\label{eq:Matter_Lagrangian-1}
\end{align}
where the covariant derivatives are defined by 
\begin{align}
\begin{aligned}D_{\mu}\phi & =\left(\nabla_{\mu}-i\Delta A_{\mu}-ia_{\mu}+i\frac{\D+1}{2}V_{\mu}\right)\phi\ ,\\
D_{\mu}\psi & =\left(\nabla_{\mu}-i(\D-1)A_{\mu}-ia_{\mu}+i\frac{\D+1}{2}V_{\mu}\right)\psi\ .
\end{aligned}
\end{align}
This is non-vanishing for $V_{\mu}=0$ and the gauge multiplet configuration
above. Hence, there are no additional zero modes coming from the dynamical
chiral multiplets.%
\footnote{Actually, this is only true for $0<\Delta<2$. However, at a superconformal
fixed point, $\Delta$ coincides with the conformal dimension of $\phi$.
Unitarity then restrict a charged multiplet to have $\Delta>1/4$. %
}

Finally the measure of the integral over the zero modes is given by
the flat measure of $\sigma_{0}$ 
\begin{align}
[d\sigma]=\frac{1}{\text{Vol}(G)}\prod_{\alpha\in\mathfrak{g}}d\sigma_{\alpha}=\frac{1}{|W|\text{Vol}(T)}\prod_{i=1}^{\text{rank}\, G}d(\sigma_{0})_{i}\,\prod_{\alpha>0}\alpha(\sigma_{0})^{2}\ ,
\end{align}
where $(\sigma_{0})_{i}$ are the Cartan parts of $\sigma_{0}$ and
\[
\prod_{\alpha>0}\alpha(\sigma_{0})^{2}\ ,
\]
is the Vandermonde determinant that arises when $\sigma_{0}$ is diagonalized.
$|W|$ is the order of the Weyl group $W$ 
and $\text{Vol}(T)$ is the volume of the maximal torus.

\subsection{\label{sub:Fluctuation-determinants}Fluctuation determinants}

A straightforward way of evaluating the equivariant Euler class appearing
in the localization formula is to compute the fluctuation (super)determinant
of the dynamical fields around the space of zero modes. This is done
by solving the eigenvalue equation for the quadratic part of the $Q$-exact
term. We do so below for the gauge and chiral multiplets.%
\footnote{Our convention for the Ricci scalar, which appears in the action for
a chiral multiplet, is that of \cite{Closset:2012ru}. With this convention,
$R$ on $S^{3}$ is negative.%
}

\subsubsection{\label{sub:Matter-one-loop-determinant}Matter one-loop determinant}

We will calculate the one-loop determinant of the chiral multiplet
around the background $V_{\mu}=0$ and $A_{\mu}$ given by (\ref{eq:smooth_SUGRA_background}).
The quadratic parts of the Lagrangian for the scalars and spinors in
\eqref{eq:Matter_Lagrangian-1} consist of the following operators
\begin{align}
\begin{aligned}\D_{\phi} & =-\nabla^{2}+\D^{2}A_{\mu}A^{\mu}+((\D-1)H+\sigma_{0})^{2}-H^{2}-\frac{\D}{4}(R-6H^2)\ ,\\
\D_{\psi} & =-i\gamma^{\mu}\left(\nabla_{\mu}-i(\D-1)A_{\mu}\right)-i\left((\D-\frac{1}{2})H+\sigma_{0}\right)\ .
\end{aligned}
\end{align}
where we must use the Ricci scalar calculated in appendix \ref{sec:The-resolved-space}.

We begin with the eigenvalue problem of the scalar Laplacian 
\begin{align}
\D_{\phi}\phi=\lambda_{s}\phi\ .
\end{align}
We Fourier decompose the scalar field as 
\begin{align}
\phi(\theta,\tau,\phi)=e^{i(m\tau+n\phi)}\phi(\theta)\ ,\qquad m,n\in\mathbb{Z}\ ,
\end{align}
which leads to the second order ordinary differential equation 
\begin{align}
\begin{aligned}\phi''(\theta)+\left(2\cot(2\theta)+\frac{f'(\theta)}{2f(\theta)}\right)\phi'(\theta)+\left(\Lambda_{1}-\frac{\Lambda_{2}^{2}}{\sin^{2}\theta}-\frac{\Lambda_{3}^{2}}{\cos^{2}\theta}\right)\phi(\theta)=0\ ,\end{aligned}
\label{ScSq}
\end{align}
where 
\begin{align}
\begin{aligned}\Lambda_{1} & \equiv\frac{\lambda_{s}}{f(\theta)}-1-\frac{(\sigma_{0}-i(\D-1)\sqrt{f(\theta)})^{2}}{f(\theta)}+\frac{f'(\theta)}{2f(\theta)}\D\cot(2\theta)\ ,\\
\Lambda_{2} & \equiv\frac{2m-\D(q\sqrt{f(\theta)}-1)}{2q}\ ,\\
\Lambda_{3} & \equiv\frac{2n-\D(p\sqrt{f(\theta)}-1)}{2p}\ .
\end{aligned}
\label{ScLambda1}
\end{align}

The eigenvalue problem for the Dirac operator acting on the spinor is 
\begin{align}
\D_{\psi}\psi=\lambda_{f}\psi\ .
\end{align}
We decompose $\psi$ as 
\begin{align}
\psi(\theta,\tau,\phi)=e^{i(m\tau+n\phi)}\left(\begin{array}{c}
\psi_{1}(\theta)\\
e^{i(\tau+\phi)}\psi_{2}(\theta)
\end{array}\right)\ ,\qquad m,n\in\mathbb{Z}\ ,
\end{align}
in the basis defined by the vielbein (\ref{eq:Resolved_space_vielbein}).
This results in a set of coupled equations 
\begin{align}
\begin{aligned}\psi_{1}'(\theta) & +\left(\D\cot(2\theta)+c_{1}\tan\theta-c_{2}\cot\theta\right)\psi_{1}(\theta)+\left(\frac{\lambda_{f}+i\sigma_{0}}{\sqrt{f(\theta)}}+c_{1}+c_{2}\right)\psi_{2}(\theta)=0\ ,\\
\psi_{2}'(\theta) & +\left((2-\D)\cot(2\theta)-c_{1}\tan\theta+c_{2}\cot\theta\right)\psi_{2}(\theta)+\left(-\frac{\lambda_{f}+i\sigma_{0}}{\sqrt{f(\theta)}}+2(1-\D)+c_{1}+c_{2}\right)\psi_{1}(\theta)=0\ ,
\end{aligned}
\end{align}
where 
\begin{align}
c_{1}=\frac{2n+\D}{2p\sqrt{f(\theta)}}\ ,\qquad c_{2}=\frac{2m+\D}{2q\sqrt{f(\theta)}}\ .\label{C1C2}
\end{align}
Solving for $\psi_{2}$ yields the second order differential equation
for $\psi_{1}$ 
\begin{align}
\begin{aligned}\psi_{1}''(\theta)+\left(2\cot(2\theta)+\frac{f'(\theta)}{2f(\theta)}\right)\psi_{1}'(\theta)+\left(\tilde{\Lambda}_{1}-\frac{\Lambda_{2}^{2}}{\sin^{2}\theta}-\frac{\Lambda_{3}^{2}}{\cos^{2}\theta}\right)\psi_{1}(\theta)=0\ ,\end{aligned}
\label{SpSq}
\end{align}
where 
\begin{align}
\begin{aligned}\tilde{\Lambda}_{1} & \equiv\frac{(\lambda_{f}+(\D-1)\sqrt{f(\theta)}+i\sigma_{0})^{2}}{f(\theta)}-1+\frac{f'(\theta)}{2f(\theta)}\D\cot(2\theta)\ .\end{aligned}
\label{SpLambda1}
\end{align}
The second order differential equations for the scalar and spinor
(\ref{ScSq}) and (\ref{SpSq}) have the same form up to the difference
of $\Lambda_{1}$ and $\tilde{\Lambda}_{1}$. To zeroth order in $\epsilon$ they can be
equal by identifying $\Lambda_{1}=\tilde{\Lambda}_{1}$ and $f\left(\theta\right)\equiv1$. Comparing
(\ref{ScLambda1}) and (\ref{SpLambda1}), we obtain 
\begin{align}
\lambda_{s}=\lambda_{f}(\lambda_{f}+2(\D-1)+2i\sigma_{0})\ .
\end{align}
This relation between the eigenvalues of the scalar and spinor gives
rise to the cancellation in the one-loop determinant of the matter
sector: given
$\lambda_{s}$, there are two solutions of the spinor eigenfunction
$\lambda_{f}^{(\pm)}$ satisfying $\lambda_{f}^{(+)}\lambda_{f}^{(-)}=\lambda_{s}$
as long as $\psi_{1}\neq0$ and $\psi_{2}\neq0$. Alternatively, we
could view this cancellation as happening in the singular space in
section \ref{sub:A-singular-supersymmetric} whose supersymmetry preserving
boundary conditions we are after.

The only contributions to the one-loop determinant thus come from
the modes with either $\psi_{1}\neq 0,\psi_{2}=0$ or $\psi_{1}=0,\psi_{2}\neq0$.
When $\psi_{1}\neq0,\psi_{2}=0$, the eigenvalue of $\lambda_{f}$
has to be (again to zeroth order in $\epsilon$) 
\begin{align}
\lambda_{f}^{(1)}=\frac{n}{p}+\frac{m}{q}-i\sigma_{0}+\frac{\D}{2}\left(\frac{1}{p}+\frac{1}{q}\right)-2(\D-1)\ ,
\end{align}
and we must solve the equation 
\begin{align}
\psi_{1}'(\theta) & +\left(\D\cot(2\theta)+\frac{2n+\D}{2p\sqrt{f(\theta)}}\tan\theta-\frac{2m+\D}{2q\sqrt{f(\theta)}}\cot\theta\right)\psi_{1}(\theta)=0\ .
\end{align}
To obtain the boundary condition at $\theta=0$ and $\theta=\pi/2$,
we solve it near the singularities 
\begin{align}
\psi_{1}(\theta)=\left\{ \begin{array}{ll}
\sin^{m}\theta+\cdots\ , & \qquad\theta=0\ ,\\
\cos^{n}\theta+\cdots\ , & \qquad\theta=\pi/2\ .
\end{array}\right.
\end{align}
We need to impose $m>-1$ and $n>-1$ for the solution to be integrable
$\int d\theta\sin\theta\cos\theta|\psi(\theta)|^{2}<\infty$.

When $\psi_{1}=0,\psi_{2}\neq0$, the eigenvalue of $\lambda_{f}$
has to be 
\begin{align}
\lambda_{f}^{(2)}=-\frac{n}{p}-\frac{m}{q}-i\sigma_{0}-\frac{\D}{2}\left(\frac{1}{p}+\frac{1}{q}\right)\ ,
\end{align}
and we solve the equation 
\begin{align}
\psi_{2}'(\theta) & +\left((2-\D)\cot(2\theta)-\frac{2n+\D}{2p\sqrt{f(\theta)}}\tan\theta+\frac{2m+\D}{2q\sqrt{f(\theta)}}\cot\theta\right)\psi_{2}(\theta)=0\ .
\end{align}
To obtain the boundary condition at $\theta=0$ and $\theta=\pi/2$,
we solve it near the singularities 
\begin{align}
\psi_{2}(\theta)=\left\{ \begin{array}{ll}
\sin^{-m-1}\theta+\cdots\ , & \qquad\theta=0\ ,\\
\cos^{-n-1}\theta+\cdots\ , & \qquad\theta=\pi/2\ ,
\end{array}\right.
\end{align}
and we need to impose $m<0$ and $n<0$ for the solution to be integrable.

In total, once we take the $\epsilon\rightarrow0$ limit, there is complete cancellation between $\lambda_{s}$ and
$\lambda_{f}$ (and $\lambda_{f}+2(\D-1)+2i\sigma_{0}$) for the modes
with $\psi_{1}\neq0,\psi_{2}\neq0$. The modes of $\lambda_{f}^{(2)}$
with $\psi_{1}=0,\psi_{2}\neq0$ have no bosonic counterparts, while
the modes of $\lambda_{f}^{(1)}$ with $\psi_{1}\neq0,\psi_{2}=0$
do not have pairing modes of $\lambda_{f}+2(\D-1)+2i\sigma_{0}$.
Taking this into account, we obtain the one-loop partition function
of the matter sector 
\begin{align}
\begin{aligned}Z_{\text{matter}}^{\text{1-loop}} & =\frac{\det\D_{\psi}}{\det\D_{\phi}}=\prod\frac{\lambda_{f}^{(2)}}{\lambda_{f}^{(1)}+2(\D-1)+2i\sigma_{0}}\ ,\\
 & =\prod_{m,n\ge0}\frac{\frac{n}{p}+\frac{m}{q}-i\sigma_{0}-\frac{\D-2}{2}\left(\frac{1}{p}+\frac{1}{q}\right)}{\frac{n}{p}+\frac{m}{q}+i\sigma_{0}+\frac{\D}{2}\left(\frac{1}{p}+\frac{1}{q}\right)}\ ,\\
 & =\Gamma_{h}\left(-\sigma_{0}+\frac{i\D}{2}\left(\frac{1}{p}+\frac{1}{q}\right);\frac{i}{p},\frac{i}{q}\right)\ ,
\end{aligned}
\label{eq:matter_one_loop}
\end{align}
where the final result is expressed in terms of the hyperbolic gamma
function defined in appendix \ref{sec:Special-functions}.

\subsubsection{\label{sub:Gauge-one-loop-determinant}Gauge one-loop determinant}

The quadratic part of the Yang-Mills Lagrangian \eqref{eq:YMaction}
is 
\begin{align}
\begin{aligned}\CL_{\text{fluc}} & =\text{Tr}\Big[a_{\mu}\D_{v}a^{\mu}-[a_{\mu},\sigma_{0}]^{2}+\partial_{\mu}\sigma'\partial^{\mu}\sigma'+(D'+\sigma')^{2}\\
 & \qquad-i\tilde{\lambda}\gamma^{\mu}(\nabla_{\mu}+iA_{\mu})\lambda-i\tilde{\lambda}[\sigma_{0},\lambda]+\frac{1}{2}\tilde{\lambda}\lambda+\partial_{\mu}\bar{c}\partial^{\mu}c\Big]\ .
\end{aligned}
\end{align}
The D-term fluctuation $D'$ can be integrated out easily. The vector
Laplacian can be written as 
\[
\D_{v}=\star\, d\star d+d\star d\,\star\ ,
\]
and we integrate out $b$ in (\ref{eq:Gauge_fixing_term}) to impose
the covariant gauge $d\star a=0$.

The vector potential one-form can be decomposed as 
\begin{align}
a=B+d\chi\ ,
\end{align}
where $B$ is a divergence-less one-form 
\[
d\star B=0\ .
\]
The covariant gauge guarantees that $\chi$ does not couple to $\sigma_{0}$
\[
\star d\star d\chi=\Delta_{\text{scalar}}\chi=0\ .
\]
The determinants from the path integral over $\chi$ and $\sigma'$
then cancel with the ghosts $c,\bar c$.

The remaining gauge fixed action for the divergence-less vectors and
the gauginos becomes 
\begin{align}
\begin{aligned}\CL_{\text{fluc}} & =\text{Tr}\Big[B_{\mu}\D_{v}B^{\mu}-[B_{\mu},\sigma_{0}]^{2}-i\tilde{\lambda}\gamma^{\mu}(\nabla_{\mu}+iA_{\mu})\lambda-i\tilde{\lambda}[\sigma_{0},\lambda]+\frac{1}{2}\tilde{\lambda}\lambda\Big]\ .\end{aligned}
\end{align}
We decompose all adjoint fields with respect to the Cartan-Weyl basis
\begin{align}
\begin{aligned}\CL_{\text{fluc}} & =\sum_{i}\left[B_{\mu}^{i}\D_{v}B_{i}^{\mu}+\tilde{\lambda}_{i}\D_{\psi}\lambda_{i}|_{\D=0,\sigma_{0}=0}\right]\\
 & \qquad+\sum_{\alpha}\left[B_{\mu}^{-\alpha}(\D_{v}+\alpha(\sigma_{0})^{2})B_{\alpha}^{\mu}+\tilde{\lambda}_{-\alpha}(\D_{\psi}|_{\D=0,\sigma_{0}=0}-i\alpha(\sigma_{0}))\lambda_{\alpha}\right]\ ,
\end{aligned}
\end{align}
where we now assume that $\sigma_{0}$ is in the Cartan subalgebra.
We already
know the spectrum of the Dirac operator for the adjoint fermion.
We need to determine the spectrum of the vector Laplacian $\D_{v}$.

Consider the eigenvalue problem of the form as in \cite{Klebanov:2011uf}
\begin{align}
d\star B=0\ ,\qquad\star dB=\lambda_{v}B\ .\label{DLVec}
\end{align}
From the explicit form of the vector Laplacian, the solution of the
eigenvalue problem yields the eigenfunction of the vector Laplacian
\begin{align}
\D_{v}B=\lambda_{v}^{2}B\ .
\end{align}
To solve (\ref{DLVec}), we expand $B$ in terms of the left-invariant
one-forms on the resolved space (\ref{eq:Resolved_space_vielbein})
\begin{align}
B=e^{i(m\tau+n\phi)}\left[e^{-i(\phi+\tau)}b_{+}(\theta)e^{+}+b_{0}(\theta)e^{3}+e^{i(\phi+\tau)}b_{-}(\theta)e^{-}\right]\ ,
\end{align}
where we defined $e^{\pm}=e^{1}\pm ie^{2}$. Then we obtain four ordinary
differential equations which can be solved by 
\begin{align}
\begin{aligned}b_{\pm}(\theta) & =\left(\frac{m}{q}+\frac{n}{p}\pm\lambda_{v}\right)\left[\pm\frac{\sqrt{f(\theta)}B'(\theta)}{2}+\frac{B(\theta)}{2}\left(\frac{m}{q}\cot\theta-\frac{n}{p}\tan\theta\right)\right]\ ,\\
b_{0}(\theta) & =\left(\frac{m}{q}+\frac{n}{p}+\lambda_{v}\right)\left(\frac{m}{q}+\frac{n}{p}-\lambda_{v}\right)B(\theta)\ ,
\end{aligned}
\end{align}
where $B(\theta)$ satisfies the second order differential equation
\begin{align}
B''(\theta)+\left(2\cot(2\theta)+\frac{f'(\theta)}{2f(\theta)}\right)B'(\theta)+\frac{1}{f(\theta)}\left(\lambda_{v}^{2}-2\sqrt{f(\theta)}\lambda_{v}-\frac{m^{2}}{q^{2}\sin^{2}\theta}-\frac{n^{2}}{p^{2}\cos^{2}\theta}\right)B(\theta)=0\ .\label{VecDE}
\end{align}
The eigenfunction of the Dirac operator satisfying 
\begin{align}
\D_{\psi}\psi|_{\D=0,\sigma_{0}=0}=\lambda_{vf}\psi\ ,
\end{align}
can be obtained by solving the second order differential equation
(\ref{SpSq}) with $\D=0$ and $\sigma_{0}=0$. Comparing the two
differential equations (\ref{SpSq}) and (\ref{VecDE}), there is
a one-to-one correspondence between the eigenfunctions of the divergence-less vector field and the adjoint fermion given either $\lambda_{v}=\lambda_{vf}$
or $\lambda_{v}=2-\lambda_{vf}$ to zeroth order of $\epsilon$.
This map does not exist when one of the component of the spinor vanishes
as we saw in the matter sector. The gaugino modes which have bosonic
counterparts, but do not pair up with $2-\lambda_{vf}$ have eigenvalues
of 
\begin{align}
\lambda_{vf}^{(1)}=\frac{n}{p}+\frac{m}{q}+2\ ,
\end{align}
with $m>-1$ and $n>-1$. This would include non-normalizable
modes for the vector field and the actual ranges turn out to be $m>0$
and $n>0$. The fermionic modes which have no bosonic counterparts
have eigenvalues of 
\begin{align}
\lambda_{vf}^{(2)}=-\frac{n}{p}-\frac{m}{q}\ ,
\end{align}
with $m<0$ and $n<0$.

Taking into account the non-canceling modes, the one-loop partition
function of the gauge sector is given by 
\begin{align}
\begin{aligned}Z_{\text{gauge}}^{\text{1-loop}} & =\prod_{i=1}^{\text{rank}\, G}\frac{\lambda_{vf}^{(2)}}{2-\lambda_{vf}^{(1)}}\prod_{\alpha>0}\frac{(\lambda_{vf}^{(2)})^{2}+\alpha(\sigma_{0})^{2}}{(2-\lambda_{vf}^{(1)})^{2}+\alpha(\sigma_{0})^{2} } \ ,\\
 & =\prod_{i=1}^{\text{rank}\, G}\frac{\prod_{m,n<0}-\frac{n}{p}-\frac{m}{q}}{\prod_{m,n>0}-\frac{n}{p}-\frac{m}{q}}\prod_{\alpha>0}\frac{\prod_{m,n<0}\left(-\frac{n}{p}-\frac{m}{q}\right)^{2}+\alpha(\sigma_{0})^{2}}{\prod_{m,n>0}\left(-\frac{n}{p}-\frac{m}{q}\right)^{2}+\alpha(\sigma_{0})^{2}}\ ,\\
 & =\sqrt{\prod_{i}^{\text{rank}\, G}\lim_{\alpha\to0}\frac{1}{\alpha(\sigma_{0})^{2}}\frac{1}{\Gamma_{h}\left(\alpha(\sigma_{0});\frac{i}{p},\frac{i}{q}\right)\Gamma_{h}\left(-\alpha(\sigma_{0});\frac{i}{p},\frac{i}{q}\right)}}\\
 & \qquad\quad\cdot\prod_{\alpha>0}\frac{1}{\alpha(\sigma_{0})^{2}}\frac{1}{\Gamma_{h}\left(\alpha(\sigma_{0});\frac{i}{p},\frac{i}{q}\right)\Gamma_{h}\left(-\alpha(\sigma_{0});\frac{i}{p},\frac{i}{q}\right)}\ ,\\
 & =(2\pi\sqrt{pq})^{\text{rank}\, G}\prod_{\alpha>0}\frac{1}{\alpha(\sigma_{0})^{2}}\frac{1}{\Gamma_{h}\left(\alpha(\sigma_{0});\frac{i}{p},\frac{i}{q}\right)\Gamma_{h}\left(-\alpha(\sigma_{0});\frac{i}{p},\frac{i}{q}\right)}\ .
\end{aligned}
\label{eq:gauge_one_loop}
\end{align}
where the result has again been expressed in terms of the hyperbolic
gamma function.

\subsection{\label{sub:Classical-contributions}Classical contributions}

To complete the localization calculation, we must evaluate the classical
action on the space of zero modes found in section \ref{sub:Zero-modes}.
As in the $S^{3}$ calculation, only the Chern-Simons and Fayet-Iliopoulos (FI)
terms give these classical contributions. Their action is given by
\begin{align}
\mathcal{L}_{\text{cl}}=\frac{k}{4\pi}\text{Tr}\left[i\varepsilon^{\mu\nu\rho}\left(a_{\mu}\partial_{\nu}a_{\rho}+\frac{2i}{3}a_{\mu}a_{\nu}a_{\rho}\right)-2D\sigma+2i\tilde{\lambda}\lambda\right]+\frac{\xi}{2\pi}\,\text{Tr}\,(D-\sigma H-a_{\mu}V^{\mu})\ ,
\end{align}
where $k$ is the Chern-Simons level and $\xi$ the FI parameter.
They give rise to classical contributions of the form 
\begin{align}
Z_{\text{classical}}=\exp\left[pq\left(i\pi k\,\text{Tr}(\sigma_{0}^{2})-2\pi i\xi\,\text{Tr}(\sigma_{0})\right)\right]\ .
\end{align}

\subsection{The matrix model}

Combining the results of sections \ref{sub:Fluctuation-determinants} and \ref{sub:Classical-contributions} we obtain the result for the
partition function on the singular space \eqref{eq:two_parameter_metric}
in terms of the $\epsilon\rightarrow0$ limit of the one on the resolved
space \eqref{eq:smooth_two_parameter_metric}. The result is a familiar matrix
model. The integration is over the Lie algebra $\mathfrak{g}$ of
the gauge group $G$. At this point it is advantageous to rescale
the integration variables by a factor of $1/\sqrt{pq}$ and use the
scale invariance of the hyperbolic gamma function (see appendix \ref{sec:Special-functions})
to set 
\begin{align*}
Z_{\text{matter}}^{\text{1-loop}} & =\Gamma_{h}\left(-\sigma_{0}+\frac{i\D}{2}\left(b+b^{-1}\right);ib,\frac{i}{b}\right)\ ,\qquad b=\sqrt{\frac{q}{p}}\ .
\end{align*}
This expression coincides with the one for the squashed sphere $S_{b}^{3}$
with squashing parameter $b$ \cite{Hama:2011ea,Imamura:2011wg}.
The same applies to the expression for the fluctuation determinant of
the vector multiplet. The rescaled contribution of a Chern-Simons
term is 
\[
Z_{\text{CS}}=\exp\left(i\pi k\,\text{Tr}\left(\sigma_{0}\right)^{2}\right) \ ,
\]
which also coincides with the squashed sphere.

We may also consider FI ($\xi$) and real mass terms ($m$). Combining
the results above for an arbitrary theory we get 
\begin{equation}
Z_{\text{singular space}}\left(\{\xi_{i}\},\{m_{j}\}|p,q\right)=Z_{S_{b}^{3}}\left(\left\{ \frac{\xi_{i}}{\sqrt{pq}}\right\} ,\left\{ \frac{m_{j}}{\sqrt{pq}}\right\} \Bigg|b=\sqrt{\frac{q}{p}}\right)\ ,\label{eq:Partition_Function_Equality}
\end{equation}
and note that 
\[
Z_{S_{b}^{3}}\left(\{\xi_{i}\},\{m_{j}\}|b=1\right)=Z_{S^{3}}\left(\{\xi_{i}\},\{m_{j}\}\right)\ .
\]

The total partition function is 
\begin{align}
\begin{aligned}
Z_{\text{singular space}}\left(\{\xi_{i}\},\{m_{j}\}|p,q\right) = &\frac{1}{|W|}\int\prod_{i=1}^{\text{rank}\, G}d\sigma_{i}\, e^{\pi ik\,\text{Tr}(\sigma^{2})-2\pi i\tilde{\xi}\,\text{Tr}(\sigma)}\cdot\prod_{\alpha}\frac{1}{\Gamma_{h}\left(\alpha(\sigma)\right)}\\
 & \cdot\prod_{I}\prod_{\rho\in\CR_{I}}\Gamma_{h}\left(\tilde{\rho}(\sigma)+\tilde{m}_{I}+i\omega\D_{I}\right)\ ,
\end{aligned}
\label{eq:Partition_Function_Total}
\end{align}
where 
\[
\Gamma_{h}\left(z\right)\equiv\Gamma_{h}\left(z;i\omega_{1},i\omega_{2}\right)\ ,\qquad\omega_{1}=\sqrt{\frac{q}{p}}\ ,\qquad\omega_{2}=\sqrt{\frac{p}{q}}\ ,
\]
is the hyperbolic gamma function in appendix \ref{sec:Special-functions} and
\[
\omega=\frac{\omega_{1}+\omega_{2}}{2}\ .
\]
$k$ is the Chern-Simons level and 
\[
\tilde{\xi}=\sqrt{pq}\,\xi\ ,\qquad\tilde{m}=\sqrt{pq}\,m\ ,
\]
are the FI and real mass parameters. $I$ labels the
types of chiral multiplets and $\rho$ is a weight in a representation
$\CR_{I}$ of a gauge group $G$. $\D_{I}$ is the $R$-charge of
the scalar field in a chiral multiplet.

\subsection{\label{sub:Supersymmetric-Renyi-entropy}Supersymmetric R{\' e}nyi
entropy}

As explained in \cite{Closset:2012vg}, the partition function of
an $\mathcal{N}=2$ theory on $S^{3}$ can receive contributions from
certain non-universal terms involving the fields in the supergravity
multiplet and, possibly, background flavor symmetry vector multiplets.
One can check, however, that the real part of the free energy is universal
and an intrinsic observable of the SCFT to which the theory flows.
The partition function \eqref{eq:Partition_Function_Total} shares these
properties. This is because the field $A_{\mu}$ is real and, as on
$S^{3}$, $H=-i$. With these values, the contribution of all non-universal
terms specified in \cite{Closset:2012vg} remains purely imaginary,
while the free energy of a conformal theory on the singular space,
as the limit from the smooth resolved space, is expected to be real.

We define the super R{\' e}nyi entropy to be 
\[
S_{q}^{\text{susy}}=\frac{1}{1-q}\Re\left[\log\left(\frac{Z_{\text{singular space}}\left(\{\xi_{i}\}=\{m_{j}\}=0|1,q\right)}{\left(Z_{S^{3}}\right)^{q}}\right)\right]\ ,
\]
hence our main result is that 
\begin{align}
S_{q}^{\text{susy}}=\frac{1}{1-q}\Re\left[\log\left(\frac{Z_{S_{b}^{3}}\left(b=\sqrt{q}\right)}{\left(Z_{S^{3}}\right)^{q}}\right)\right]\ .\label{eq:SRE-1}
\end{align}

\section{\label{sec:Checks}Checks}

We have embedded the metric of the branched sphere into a supergravity
background. In order to preserve supersymmetry on this (singular)
space, we were forced to turn on an additional background $R$-symmetry
gauge field. Since this deformation is not part of the original definition
of the R{\' e}nyi entropy we should explain its appearance. Note that
the deformation is pure gauge away from the singularity and can, hence,
be traded for boundary conditions on the $R$-charged fields by using
a singular background $R$-symmetry gauge transformation.%
\footnote{This is no longer true for the resolved space.%
} The space is then the original branched sphere, except
for the effects associated with the imaginary
value of $H$, with supersymmetry preserving boundary conditions for
all fields.

\subsection{Limits and behavior}

An immediate consequence of the formula in section \ref{sub:Supersymmetric-Renyi-entropy}
is that for a topological theory the super R{\' e}nyi entropy is $q$-independent.
This is because it is numerically equal to the partition function
of a smooth space: the squashed sphere.%
\footnote{Of course the usual framing dependence of the Chern-Simons theory
applies.%
}

The behavior around $q=1$ for a general theory is such that 
\[
S_{q}^{\text{susy}}\quad\xrightarrow[q\rightarrow1]{}\quad S=-F\ ,
\]
where $S$ is the entanglement entropy and $F$ is the free energy
on $S^{3}$, which implies 
\begin{equation}
\partial_{q}Z_{\text{singular space}}\left(\{\xi_{i}\}=\{m_{j}\}=0|1,q\right)|_{q=1}=0\ ,\label{eq:Vanishing_at_q1}
\end{equation}
and this statement in turn follows from the more general observation
that 
\[
Z_{\text{singular space}}\left(\{\xi_{i}\}=\{m_{j}\}=0|1,q\right)=Z_{\text{singular space}}\left(\{\xi_{i}\}=\{m_{j}\}=0\bigg|1,\frac{1}{q}\right)\ .
\]

The next term in the expansion around $q=1$ is also interesting.
The squashed sphere partition function has been shown to have the
following expansion around $b=1$ \cite{Closset:2012ru}%
\footnote{The possibility of $\tau_{rr}$ being a $c$-function in three-dimensions
has been recently tested and ruled out by \cite{Nishioka:2013gza}.%
} 
\begin{align}
\begin{aligned}
F_{b} & =  -\log Z_{S_{b}^{3}\ ,}\\
\partial_{b}F_{b}|_{b=1} & = 0\ ,\\
\Re\left(\partial_{b}^{2}F_{b}\right)|_{b=1} & = \frac{\pi^{2}}{2}\tau_{rr}\ ,
\end{aligned}
\end{align}
where $\tau_{rr}$ is a constant which appears in the SCFT flat space
correlation functions at separated points 
\begin{align}
\begin{aligned}\langle j_{\mu}^{\left(R\right)}\left(x\right)j_{\mu}^{\left(R\right)}\left(0\right)\rangle & =\frac{\tau_{rr}}{16\pi^{2}}\left(\delta_{\mu\nu}\partial^{2}-\partial_{\mu}\partial_{\nu}\right)\frac{1}{x^{2}}\ ,\\
\langle T_{\mu\nu}\left(x\right)T_{\rho\sigma}\left(0\right)\rangle & =-\frac{\tau_{rr}}{64\pi^{2}}\left(\delta_{\mu\nu}\partial^{2}-\partial_{\mu}\partial_{\nu}\right)\left(\delta_{\rho\sigma}\partial^{2}-\partial_{\rho}\partial_{\sigma}\right)\frac{1}{x^{2}}\\
 & \quad+\frac{\tau_{rr}}{64\pi^{2}}\left(\left(\delta_{\mu\rho}\partial^{2}-\partial_{\mu}\partial_{\rho}\right)\left(\delta_{\nu\sigma}\partial^{2}-\partial_{\nu}\partial_{\sigma}\right)+\left(\mu\leftrightarrow\nu\right)\right)\frac{1}{x^{2}}\ .
\end{aligned}
\end{align}
Note that 
\[
\Re\left(\partial_{b}^{2}F_{b}\right)|_{b=1}=-\Re\left(\frac{\partial_{b}^{2}Z_{S_{b}^{3}}}{Z_{S^{3}}}\right)\Bigg|_{b=1}\ ,
\]
which implies 
\[
S_{q}^{\text{susy}}=S + \frac{\pi^{2}}{16}\tau_{rr}\left(q-1\right)+O\left(\left(q-1\right)^{2}\right)\ .
\]

\subsection{Free fields}

We can compare the results of section \ref{sec:Localization} to the previous
results for free field theories \cite{Klebanov:2011uf}. Free (complex)
scalars have canonical dimension $\Delta=1/2$. Consider the action
for a single uncharged chiral multiplet with $\Delta=1/2$ coupled
to the supergravity background of the singular space given in section
\ref{sub:A-singular-supersymmetric}. There are two major differences
between this action and the one used to compute the R{\' e}nyi entropies
of free scalars and fermions in \cite{Klebanov:2011uf} 
\begin{enumerate}
\item The supersymmetric action \eqref{eq:Matter_Lagrangian-1} includes
a coupling to a background $R$-symmetry gauge field $A_\mu$. This background
gauge field is flat away from the singularities. 
\item The action \eqref{eq:Matter_Lagrangian-1} includes a coupling of the
scalars to the Ricci scalar $R$. Excising the loop at $\theta=0$
means ignoring the delta function supported there in the expression
for $R$ given by \eqref{eq:RSdelta} taken at $p=1$. We have, instead,
regularized this contribution by smoothing out the space and then
taking an appropriate limit. 
\end{enumerate}
Both of the deformations above can
be reformulated as alternative boundary conditions for bosons and
fermions. A choice of such boundary conditions is already implicit
in the flat space calculation of the R{\' e}nyi entropy. Interestingly,
for odd integer $q$, one can still rewrite the results of \cite{Klebanov:2011uf}
adapted to a free conformally coupled chiral multiplet in a way that
resembles a supersymmetric calculation 
\begin{equation}
Z_{\text{non-susy}}\left(q\right)=\Gamma_{h}\left(\frac{i}{2};i,i\right)\prod_{r=1}^{\frac{q-1}{2}}\Gamma_{h}\left(\frac{i}{2}+i\frac{r}{q};i,i\right)^{2}\ .\label{eq:Renyi_entropy_non_susy_free_chiral}
\end{equation}

\subsection{Duality}
 
If the super R{\' e}nyi entropy is computed by embedding a 3d SCFT
into a flow from a UV action, which is used in the localization, we
expect the result to be an intrinsic (scheme independent) observable
of the IR fixed point. This is in line with the definition of the
usual R{\' e}nyi entropy as an operation on a density matrix obtained
from the ground state wave function.

One way to test these properties of the super R{\' e}nyi entropy is
to compare its value for a pair of IR dual quantum field theories.
Examples of such dualities abound in 3d gauge theories with $\mathcal{N}\ge2$
supersymmetry. These are roughly classified into 3d mirror symmetry
\cite{Intriligator:1996ex,deBoer:1996ck,deBoer:1996mp,deBoer:1997ka,Jensen:2009xh}
and Seiberg-like duality \cite{Aharony:1997gp,Giveon:2008zn,Niarchos:2008jb}.
Extensive comparisons have been made for the partition functions of
the theories involved: for $S^{3}$ in \cite{Kapustin:2010mh,Kapustin:2010xq,Willett:2011gp};
the squashed sphere, $S_{b}^{3}$, in \cite{Benini:2011mf,Yaakov:2013fza};
for $S^{2}\times S^{1}$ (the superconformal index) in \cite{Imamura:2011su,Kim:2013cma,Bashkirov:2011vy,Bashkirov:2011pt,Kapustin:2011jm,Krattenthaler:2011da,Kapustin:2011vz,Hwang:2011ht,Hwang:2011qt}.

The results in section \ref{sec:Localization} relate the super R{\' e}nyi
entropy to the $S_{b}^{3}$ partition function. This makes such a duality
comparison for the super R{\' e}nyi entropy, or equivalently the branched
sphere partition function, fairly trivial. We note that, as in the
case of other partition functions, we can further deform the computation
by real mass and FI terms and compare the result as a function of
them. Equation \eqref{eq:Partition_Function_Equality} guarantees
that the deformed expression is duality invariant. Note that one could
naively use the expression \eqref{eq:Renyi_entropy_non_susy_free_chiral}
as the one-loop determinant for a chiral superfield. Such an approach
does not lead to a duality invariant expression.

\subsection{The defect operator interpretation}

The R{\' e}nyi entropy associated with a region $V$ at integer $q$
can be computed using the replica trick. This involves defining the
theory on a $q$-covering space with conical singularities. If the
theory in question is free, there is an alternative formulation of
this procedure \cite{Casini:2009sr}. One first splits each field
$\Phi$ on the covering space into $q$ fields 
\begin{equation}
\{\Phi_{n}\}_{n=1}^{q}\ ,\label{eq:Field_Splitting}
\end{equation}
each defined on a single sheet. The R{\' e}nyi entropy
is computed by the partition function of this theory with boundary
conditions such that in crossing $V$ the vector 
\[
\overrightarrow{\Phi}=\left(\begin{array}{c}
\Phi_{1}\\
\Phi_{2}\\
\vdots\\
\Phi_{q}
\end{array}\right)\ ,
\]
gets multiplied by the matrices $T$ or $T^{-1}$ given by (take $\pm1$
for bosons and fermions respectively) 
\[
T=\left(\begin{array}{ccccc}
0 & 1\\
 & 0 & 1\\
 &  & \ddots & \ddots\\
 &  &  & 0 & 1\\
\left(\pm1\right)^{q+1} &  &  &  & 0
\end{array}\right)\ .
\]
Equivalently, 
\begin{equation}
\Phi_{n}\mid_{V_{+}}=\Phi_{n+1}\mid_{V_{-}}\ ,\label{eq:Product_Theory_Boundary_Conditions}
\end{equation}
where $V_{\pm}$ denote the two ``sides'' of the $(d-1)$-dimensional
region $V$, whose boundary $\partial V$ is the ``entangling surface'',
and the condition relates $n=q$ to $n=1$ (possibly with a sign).
This condition is diagonalized by fields $\tilde{\Phi}_{n}$ with
monodromy 
\begin{equation}
\beta_{n}=e^{2\pi i\frac{n}{q}}\ , \qquad \begin{cases}
n\in\left\{ 0,\cdots,q-1\right\}\ ,  & \text{bosons}\ ,\\
n\in\left\{ -\frac{q-1}{2},\cdots,\frac{q-1}{2}\right\}\ ,  & \text{fermions}\ ,
\end{cases}\label{eq:defect_monodromies}
\end{equation}
around the singularity at $\partial V$.

A different interpretation of the procedure above is the evaluation
(in the free theory) of the product of expectation values of certain
defect operators. These defects are supported on $\partial V$ and
are defined to reproduce the monodromies $\beta_{n}$. In a 2d CFT
these would be twist operators defined at the boundary of the entangling
interval. In a 3d theory, we expect line (or loop) operators. Since
the computation of the expectation value of a $Q$-closed observable
in a supersymmetric theory using localization is essentially (ignoring
the moduli) equivalent to a free field computation, we might expect
that a similar result holds for the super R{\' e}nyi entropy.

Consider the subset of the partition functions in section \ref{sec:Localization}
for parameters $p=1,q\in\mathbb{N}$. We can rewrite the fluctuation
determinant \eqref{eq:matter_one_loop} of a chiral field with the
help of the identity \eqref{HGidentity} 
\begin{align}
\begin{aligned}Z_{\text{matter}}^{\text{1-loop}}\left(\sigma,\Delta,q\right) & =\Gamma_{h}\left(-\sigma_{0}+\frac{i\Delta}{2}\left(1+\frac{1}{q}\right);i,\frac{i}{q}\right)\ ,\\
 & =\prod_{r=0}^{q-1}\Gamma_{h}\left(-\sigma_{0}+i\left(\frac{\Delta}{2}\left(1+\frac{1}{q}\right)+\frac{r}{q}\right);i,i\right)\ ,\\
 & =\prod_{r=0}^{q-1}Z_{\text{matter}}^{\text{1-loop}}\left(-\sigma_{0}+i\left(\frac{\Delta}{2}\left(\frac{1}{q}-1\right)+\frac{r}{q}\right),\Delta,1\right)\ .
\end{aligned}
\end{align}
The terms appearing in the final product can be reinterpreted as coming
from the introduction of a supersymmetric abelian vortex loop of charge
\[
q_{r}^{\text{vortex}}=\frac{\Delta}{2}\left(\frac{1}{q}-1\right)+\frac{r}{q}\ ,
\]
supported on the great circle $\theta=0$. These affect the fluctuation
determinant of a chiral multiplet on the round $S^{3}$ in exactly
this way \cite{Kapustin:2012iw,Drukker:2012sr}.

When considering the gauge sector we must deal with the moduli. We
do not expect to be able to express the contribution of these modes
as the determinant arising from a free theory in the presence of a
defect. Rather, we should try to apply the splitting \eqref{eq:Field_Splitting}
and the boundary conditions \eqref{eq:Product_Theory_Boundary_Conditions}
directly. The boundary conditions (\ref{eq:Product_Theory_Boundary_Conditions})
imply that the moduli defined on different sheets, which we denote
by $a_{n}=\left(\sigma_{0}\right)_{n}$, are related by 
\begin{equation}
a_{n}=a_{n+1}\ .\label{eq:Moduli_Boundary_Conditions}
\end{equation}
There are therefore $q-1$ such ``delta function'' insertions in
the matrix model. This would be somewhat ill-defined if the $a_{n}$
each transformed in the adjoint representation of a different copy
of the gauge group. We therefore consider a single gauge group with
$q$ independent adjoint valued fields $\sigma_{n}$ of which the
$a_{n}$ are the constant modes. We expect the part of the matrix
model coming from the gauge multiplets to be of the form 
\[
\frac{1}{\text{Vol}\left(G\right)}\int\left(\prod_{m=1}^{q-1}\delta\left(a_{m}-a_{m+1}\right)\right)\prod_{r=0}^{q-1}Z_{\text{vector}}^{\text{1-loop}}\left(a_{r}|q_{r}^{\text{vortex}}\right)da_{r}\ ,
\]
where, following either the expansion in section \ref{sub:Gauge-one-loop-determinant}
or the evaluation in \cite{Drukker:2012sr} 
\begin{align}
\begin{aligned}Z_{\text{vector}}^{\text{1-loop}} & \left(a|q^{\text{vortex}}\right)\\
= & \left\{ \begin{array}{ll}
\left(2\pi\right)^{\text{rank}\, G}\prod_{\alpha>0}\frac{1}{\alpha(a)^{2}}\frac{1}{\Gamma_{h}\left(\alpha(a);i,i\right)\Gamma_{h}\left(-\alpha(a);i,i\right)}\ , & q^{\text{vortex}}=0\ ,\\
\Gamma_{h}\left(iq^{\text{vortex}};i,i\right)^{-\text{rank}\, G}\prod_{\alpha>0}\frac{1}{\Gamma_{h}\left(\alpha(a)+iq^{\text{vortex}};i,i\right)\Gamma_{h}\left(-\alpha(a)+iq^{\text{vortex}};i,i\right)}\ , & q^{\text{vortex}}\ne0\ .
\end{array}\right.
\end{aligned}
\end{align}
We set
\[
q_{r}^{\text{vortex}}=\frac{r}{q}\ .
\]
Using the identities
\[
\prod_{r=1}^{q-1}\frac{1}{\Gamma_{h}\left(i\frac{r}{q};i,i\right)}=\sqrt{q}\ ,
\]
and \eqref{HGidentity} we recover the one-loop determinant of the gauge sector \eqref{eq:gauge_one_loop} 
\begin{align}
\begin{aligned}\frac{1}{\text{Vol}\left(G\right)}\int\left(\prod_{m=1}^{q-1}\delta\left(a_{m}-a_{m+1}\right)\right)\prod_{r=0}^{q-1} & Z_{\text{vector}}^{\text{1-loop}}\left(a_{r}|q_{r}^{\text{vortex}}\right)da_{r}\\
 & \rightarrow\quad\frac{1}{\text{Vol}\left(G\right)}\int Z_{\text{vector}}^{\text{1-loop}}\left(a,q\right)da\ .
\end{aligned}
\end{align}
The contribution of a classical Chern-Simons term is similarly 
\[
\exp\left(i\pi k\sum_{n=1}^{q}\text{Tr}\left(a_{n}\right)^{2}\right)\quad\rightarrow\quad\exp\left(i\pi qk\,\text{Tr}\left(a\right)^{2}\right)\ .
\]

The upshot is that the supersymmetric R{\' e}nyi entropy can indeed
be re-expressed in terms of (supersymmetric) defect operators.\footnote{The physical defects are in the gravitational and $R$-symmetry backgrounds.
We have re-expressed their effect as (fictitious) flavor/gauge defects
acting on free fields. %
}
Note that the defect operator charge for a chiral superfield is not simply
the monodromy implied by \eqref{eq:defect_monodromies}. The extra
$\Delta$ dependent charge reflects the partial twisting inherent
in the definition of the supersymmetric R{\' e}nyi entropy.

\section{\label{sec:Examples}Examples}

In this section, we present several examples to show how to compute
the super R{\' e}nyi entropy and show how it behaves as a function
of $q$. To this end, we rewrite the super R{\' e}nyi entropy \eqref{eq:SRE-1}
in terms of the free energy $F(q)\equiv-\log |Z_\text{singular space}(1,q)|$ on the branched
sphere 
\begin{align}
S_{q}^{\text{susy}}=\frac{1}{1-q}(qF(1)-F(q))\ .\label{SRenyi}
\end{align}
The computation is carried out in two steps: 
\begin{itemize}
\item compute the value of $R$-charge $\D$ where the free energy $F(1)$
on the round sphere is extremized, 
\item evaluate \eqref{SRenyi} with the obtained $R$-charge. 
\end{itemize}
We are mostly interested in the limits $q\to 0$, $q\to 1$ where
\begin{align}
\begin{aligned}
S_{q}^{\text{susy}}\quad \xrightarrow[q\to 0]{} \quad - F(0) \ ,\\
S_{q}^{\text{susy}}\quad \xrightarrow[q\to 1]{} \quad - F(1) \ ,\\
\end{aligned}
\end{align}
and $q\to \infty$.
Analytic evaluation of \eqref{SRenyi} is still
difficult. We will mostly
compute it numerically except in the large-$N$ limit. 
We will show, for instance, that  the super R{\' e}nyi entropy is a monotonically
decreasing function of $q$.

\subsection{$\CN=4$ SQED with one flavor}

$\CN=4$ SQED with one hypermultiplet flavor is dual to the free theory
of a twisted hypermultiplet \cite{Kapustin:1999ha}. In terms of an
$\CN=2$ theory, an $\CN=4$ hypermultiplet is a pair of $\CN=2$
chiral multiplets with $R$-charge $1/2$ and in conjugate representations
of the gauge group.

We first consider a free chiral multiplet of $R$-charge $\D$ whose
partition function is given by 
\begin{align}
\begin{aligned}F_{\text{chiral}}(q,\D) & =-\log\Gamma_{h}(i\omega\D)\ ,\\
 & =\int_{0}^{\infty}\frac{dx}{2x}\left(\frac{2(1-\D)\omega}{x}-\frac{\sinh(2(1-\D)\omega x)}{\sinh(bx)\sinh(b^{-1}x)}\right)\ ,
\end{aligned}
\end{align}
where $b=\sqrt{q}$ and then $\omega=(b+b^{-1})/2$. 
For $q=1$ and $\D =1/2$, the integral can be performed
\begin{align}
F_\text{chiral}\left(1,\frac{1}{2}\right) = \frac{\log 2}{2} \ ,
\end{align}
and 
\begin{align}\label{eq:S1SQED}
S_{1}^{\CN=4\, \text{SQED}}= - \log 2 \ .
\end{align}
In the $q\to\infty$ limit
\begin{align}\label{qinfty}
\begin{aligned}F_{\text{chiral}}(q,\D)\quad \xrightarrow[q\to\infty]{}\quad & ~q\int_{0}^{\infty}\frac{dx}{2x^{2}}\left(1-\D-\frac{\sinh((1-\D)x)}{\sinh(x)}\right)\ ,\\
 & =\frac{iq}{4\pi}\left(\text{Li}_{2}(e^{-\pi i\D})-\text{Li}_{2}(e^{\pi i\D})\right)\ ,
\end{aligned}
\end{align}
and the super R{\' e}nyi entropy becomes 
\begin{align}
S_{q}^{\CN=4\, \text{SQED}}\quad \xrightarrow[q\to\infty]{}\quad \frac{G}{\pi}-\log2\simeq-0.402\ ,\label{eq:SRNN4QCD}
\end{align}
where $G$ is the Catalan's constant.
Since the free energy is invariant under $q\to 1/q$, the entropy near $q=0$ behaves as
\begin{align}
S_{q}^{\CN=4\, \text{SQED}}\quad \xrightarrow[q\to0]{}\quad -  \frac{G}{\pi q} \ .
\end{align}

\begin{figure}[htbp]
\centering 
\includegraphics[width=9cm]{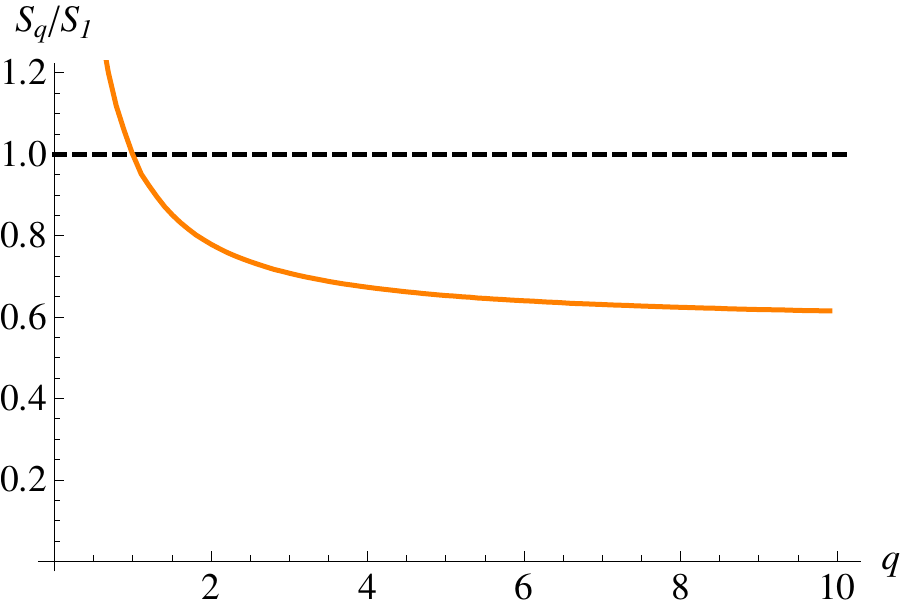}
\caption{The super R{\' e}nyi entropy of $\CN=4$ SQED with one hypermultiplet.}
\label{fig:SRHypRatio} 
\end{figure}

We compute the super R{\' e}nyi entropy of generic $q$ numerically.
The plot of the ratio between $S_q$ and $S_1$ is shown in figure \ref{fig:SRHypRatio}. It becomes $-F(1)=-\log2$
for $q=1$ and monotonically decreases and asymptotes to the value
of (\ref{eq:SRNN4QCD}) divided by \eqref{eq:S1SQED} in the large-$q$ limit. 

\subsection{$\CN=2$ SQED with one flavor}

\begin{figure}[htbp]
\centering 
\includegraphics[width=9cm]{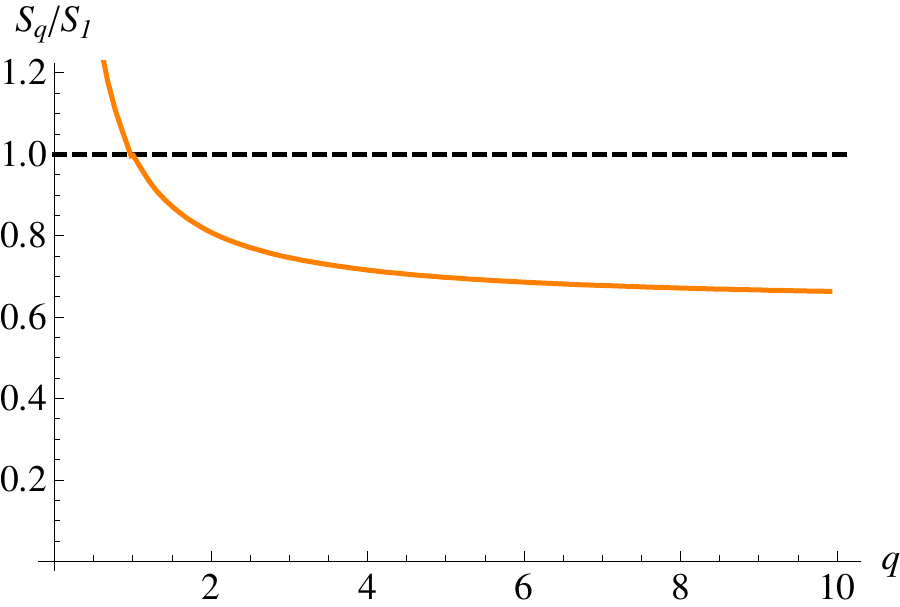}
\caption{The super R{\' e}nyi entropy of $\CN=2$ SQED with two chiral multiplet.}
\label{fig:SRTwoChRatio} 
\end{figure}

We consider $\CN=2$ SQED with two chiral multiplets of opposite charges.
This is dual to the $XYZ$ model with $R$-charges $\D=2/3$ at the
IR fixed point \cite{Aharony:1997bx}. The super R{\' e}nyi entropy at $q=1$ is minus the
free energy on a round three-sphere
\begin{align}
S_{1}^{\CN=2\, \text{SQED}}=-F^\text{XYZ}(1)=-3F_{\text{chiral}}(1,2/3)\simeq-0.872,
\end{align}
while the value in the large-$q$ limit is 
\begin{align}
S_{q}^{\CN=2\, \text{SQED}}\quad\xrightarrow[q\to\infty]{}\quad S_{1}^{\CN=2\, \text{SQED}}+\frac{3F_{\text{chiral}}(q,2/3)}{q}\simeq-0.549\ .
\end{align}
The $q\to 0$ limit follows from \eqref{qinfty}
\begin{align}
S_{q}^{\CN=2\, \text{SQED}}\quad\xrightarrow[q\to 0]{}\quad \frac{\psi
   ^{(1)}\left(\frac{2}{3}\right)-\psi
   ^{(1)}\left(\frac{1}{3}\right)}{4 \sqrt{3}
   \pi  q} \simeq - \frac{0.323}{q} \ ,
\end{align}
where $\psi^{(n)}(z)$ is the polygamma function.
The numerical plot is shown in figure \ref{fig:SRTwoChRatio}, which
asymptotes to $0.63 \sim (-0.549)/(-0.872)$ in the large-$q$ limit as expected.

A related simple example is a theory of one chiral multiplet $\Phi$ with a superpotential
\begin{align}
W=\frac{g}{6}\Phi^{3}\ .
\end{align}
The $R$-charge of $\Phi$ is canonical $\D=1/2$ in the UV, but becomes
$\D=2/3$ in the IR fixed point. The super R{\' e}nyi entropy is given
by $1/3$ of that of the $XYZ$ model in the previous section. This
theory is known to emerge on the surface of topological insulators
in four-dimensions \cite{Grover:2012bm,Lee:2006if}.

\subsection{ABJM model}

As a more elaborate example, we consider the ABJM model with $U(N)_{k}\times U(N)_{-k}$
groups \cite{Aharony:2008ug}. There are four bifundamental chiral
multiplets, two of which are in ${\bf N}\times\overline{{\bf N}}$ representation
and the other two are in $\overline{{\bf N}}\times{\bf N}$ representation.
We study the simplest case of $N=1$ where the partition function
is given by double integrals 
\begin{align}
\begin{aligned}Z_\text{ABJM}(1,k,b) & =\int_{-\infty}^{\infty}d\sigma d\tilde{\sigma}\, e^{\pi ik(\sigma^{2}-\tilde{\sigma}^{2})}\Gamma_{h}\left[\sigma-\tilde{\sigma}+\frac{i\omega}{2}\right]^{2}\Gamma_{h}\left[\tilde{\sigma}-\sigma+\frac{i\omega}{2}\right]^{2}\ ,\\
 & =\int_{-\infty}^{\infty}d\sigma d\tilde{\sigma}\, e^{\pi ik(\sigma^{2}-\tilde{\sigma}^{2})}e^{-2\int_{0}^{\infty}\frac{dx}{x}\left(\frac{\omega}{x}-\frac{\sinh(x(\omega+2i(\sigma-\tilde{\sigma}))+\sinh(x(\omega-2i(\sigma-\tilde{\sigma}))}{2\sinh(bx)\sinh(x/b)}\right)}\ ,\\
 & =\frac{1}{k}\exp\left[-2\int_{0}^{\infty}\frac{dx}{x}\left(\frac{\omega}{x}-\frac{\sinh(\omega x)}{\sinh(bx)\sinh(x/b)}\right)\right]\ ,
\end{aligned}
\end{align}
resulting in the super R{\' e}nyi entropy of the ABJM model with $N=1$
\begin{align}
S_{q}^{\text{ABJM},N=1}=-\log k-\frac{2}{1-q}\int_{0}^{\infty}\frac{dx}{x}\left(\frac{\omega-q}{x}-\frac{\sinh(\omega x)}{\sinh(bx)\sinh(x/b)}+\frac{q}{\sinh(x)}\right)\ .
\end{align}
The entanglement entropy
is obtained in $q\to1$ limit 
\begin{align}
S_{1}=-\log k-2\log2\ .
\end{align}
In the $q\to\infty$ limit, it becomes
\begin{align}\label{eq:ABJM-large-q}
\begin{aligned}
S_{q}^{\text{ABJM},N=1}\quad\xrightarrow[q\to \infty]{}\quad &-\log k - \int_0^\infty \frac{dx}{x^2}\left(
1 - \frac{2(x - \sinh (x/2))}{\sinh(x)} \right) \ ,\\
&\simeq -\log k - 0.803 \ ,
\end{aligned}
\end{align}
while it behaves near $q=0$ as
\begin{align}
S_{q}^{\text{ABJM},N=1}\quad\xrightarrow[q\to 0]{}\quad \frac{1}{q}\int_0^\infty \frac{dx}{x^2} \left( \frac{1}{\cosh(x/2)} - 1 \right) = - \frac{2G}{\pi q} \ .
\end{align}

\begin{figure}[htbp]
\centering 
\includegraphics[width=9cm]{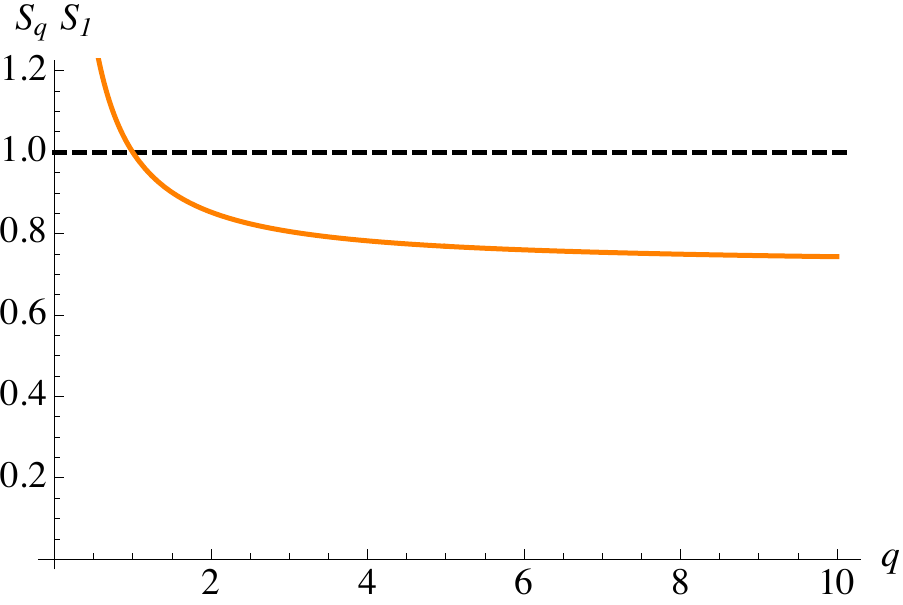} 
\caption{The super R{\' e}nyi entropy of the ABJM model of $N=1$ and $k=2$.}
\label{fig:SRE_ABJM} 
\end{figure}

The numerical plot for $k=2$ is given in figure \ref{fig:SRE_ABJM}, that is
also monotonically decreasing with respect to $q$ and approaches
to the critical value $\simeq 0.72$ in the large-$q$ limit.

\subsection{Large-$N$ limit}

\begin{figure}[htbp]
\centering 
\includegraphics[width=9cm]{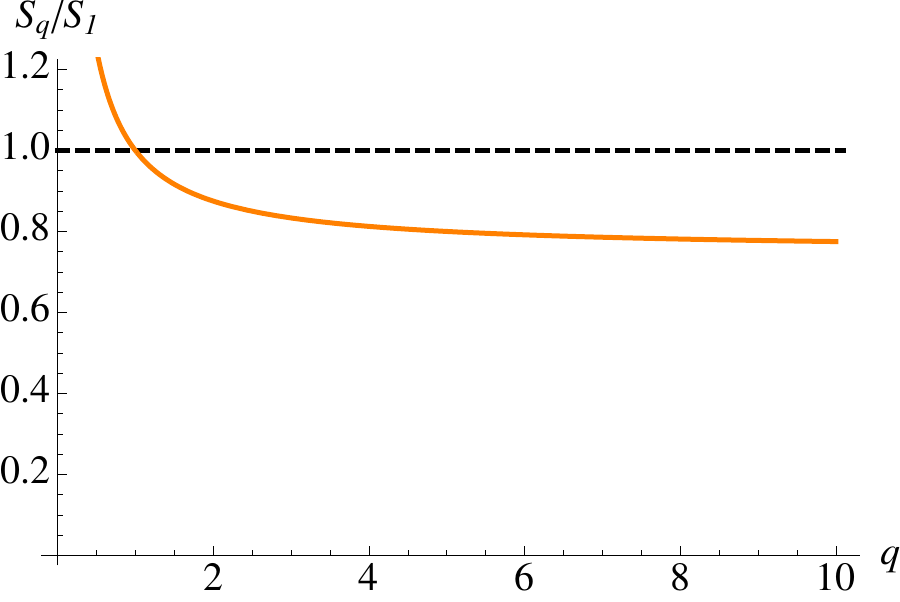} 
\caption{The super R{\' e}nyi entropy in the large-$N$ limit.}
\label{fig:SRE_LargeN} 
\end{figure}

The partition function can be solved in the large-$N$ limit as in \cite{Imamura:2011wg}
\begin{align}
F(q)=\frac{(q+1)^{2}}{4q}F(1)\ ,
\end{align}
and the super R{\' e}nyi entropy takes a simple form 
\begin{align}
S_{q}^\text{Large-$N$}=-\frac{3q+1}{4q}F(1)=\frac{3q+1}{4q}S_{1}^\text{Large-$N$}\ .\label{LargeNSRE}
\end{align}
The $q$-dependence of the super R{\' e}nyi entropy in the large-$N$
limit is shown in figure \ref{fig:SRE_LargeN}. It is monotonically
decreasing and asymptotes to $3/4$ in $q\to\infty$ limit. The
dependence on $q$ is quite similar to those of the previous sections
as well as the holographic results given in \cite{Hung:2011nu}. The
asymptotic value of $S_{q}^\text{Large-$N$}/S_{1}^\text{Large-$N$}$, however, does not agree with those
of \cite{Hung:2011nu} since their gravity solution is not supersymmetric.
It would be interesting to construct a supersymmetric gravity background
dual to our theory on the $q$-covering space by introducing a background
bulk $U(1)$ gauge field dual to the $R$-current in a similar manner
to \cite{Martelli:2011fw}.

It follows from (\ref{LargeNSRE}) that the ratio of the super R{\' e}nyi
entropy $H_{q}\equiv S_{q}^{\text{susy}}/S_{1}^{\text{susy}}$ in the large-$N$ limit satisfies
the following inequalities
\begin{align}
\begin{aligned}\partial_{q}H_{q} & \le0\ ,\\
\partial_{q}\left(\frac{q-1}{q}H_{q}\right) & \ge0\ ,\\
\partial_{q}\left((q-1)H_{q}\right) & \ge0\ ,\\
\partial_{q}^{2}\left((q-1)H_{q}\right) & \le0\ .
\end{aligned}
\end{align}
These are the same inequalities satisfied by the usual R{\' e}nyi entropy (see for instance \cite{zyczkowski2003renyi}).
We however note that our $S_{1}^\text{susy}$ is always negative. 
The sign change may come
from the renormalization of the UV divergences that could make the
R{\' e}nyi entropy negative.

\section{Conclusion}

We have defined a quantity, which we call the super R{\' e}nyi entropy,
for a 3d SCFT with $\mathcal{N}\ge2$ by considering the partition
function in a specific supergravity background. We have shown that
the super R{\' e}nyi entropy can be computed exactly using the method of
localization. This quantity differs from the R{\' e}nyi entropy of
the SCFT by a twist involving the $R$-symmetry of the theory. However,
it shares many of the properties of the usual R{\' e}nyi entropy 
\begin{itemize}
\item it is an intrinsic observable of a 3d CFT, 
\item it is independent of $q$ for topological theories, 
\item it can be used to recover the entanglement entropy, although in a
rather trivial way, 
\item it has nice analytic properties in the large-$N$ limit, 
\end{itemize}
and has additional desirable properties 
\begin{itemize}
\item deformation invariance makes it possible to calculate the super R{\' e}nyi
entropy of a SCFT arising as an IR fixed point by localizing the UV
action, 
\item the result can sometimes be written down exactly, even for strongly
coupled theories, using special functions, 
\item duality invariance can be checked explicitly,  
\item analytic continuation to non-integer $q$ is automatic. 
\end{itemize}
We have seen that the expansion of the super R{\' e}nyi entropy includes
the usual entanglement entropy and also a coefficient of the two-point
function of the energy-momentum tensors or $R$-symmetry currents. Both
quantities have been previously calculated using localization. It
would be interesting to see if additional simple SCFT observables
arise as limits.

\acknowledgments We would like to thank T.\,Dumitrescu, G.\,Festuccia,
I.\,Klebanov, R.\,Myers, B.\,Safdi, B.\,Willett and K.\,Yonekura for valuable discussions.
This research is supported by NSF grant PHY-0756966.

\appendix

\section{\label{sec:Conventions}Conventions}

We use a Clifford algebra with the Pauli matrices 
\begin{equation}
\left[\sigma_{i},\sigma_{j}\right]=2i{\varepsilon_{ij}}^{k}\sigma_{k}\ ,\qquad\left\{ \sigma_{i},\sigma_{j}\right\} =2\delta_{ij}\ ,\label{eq:Clifford_Algebra}
\end{equation}
where indices $i,j$ are raised and lowered by ${\delta^{i}}_{j}$
and 
\[
\gamma_{\mu}:={e_{\mu}}^{i}\sigma_{i}\ .
\]
Our convention is such that 
\begin{equation}
{{\omega_{\mu}}^{i}}_{j}={e_{\nu}}^{i}\nabla_{\mu}{e^{\nu}}_{j}\ ,\label{eq:Spin_Connection}
\end{equation}
and that%
\footnote{The convention in \cite{Closset:2012ru} for the spin connection differs
by a sign from \eqref{eq:Spin_Connection}. However, the covariant
derivative is also defined with an additional sign. %
}

\begin{align*}
\nabla_{\mu}\varepsilon & =\left(\partial_{\mu}+\frac{1}{8}{\omega_{\mu}}^{ij}\left[\sigma_{i},\sigma_{j}\right]\right)\varepsilon\ .
\end{align*}

\section{\label{sec:The-branched-sphere}The branched sphere}

For the branched sphere \eqref{eq:two_parameter_metric} we choose
a vielbein 
\begin{align}
\begin{aligned}e^{1} & =\sin\left(\tau+\phi\right)d\theta+q\frac{\sin2\theta}{2}\cos\left(\tau+\phi\right)d\tau-p\frac{\sin2\theta}{2}\cos\left(\tau+\phi\right)d\phi\ ,\\
e^{2} & =-\cos\left(\tau+\phi\right)d\theta+q\frac{\sin2\theta}{2}\sin\left(\tau+\phi\right)d\tau-p\frac{\sin2\theta}{2}\sin\left(\tau+\phi\right)d\phi\ ,\\
e^{3} & =q\sin^{2}\theta d\tau+p\cos^{2}\theta d\phi\ .\label{eq:vielbein}
\end{aligned}
\end{align}
with a spin connection 
\begin{align}
\begin{aligned}\omega_{~2}^{1} & =(1-q\cos^{2}\theta)d\tau+(1-p\sin^{2}\theta)d\phi\ ,\\
\omega_{~3}^{1} & =\cos(\tau+\phi)d\theta-\sin\theta\cos\theta\sin(\tau+\phi)(qd\tau-pd\phi)\ ,\\
\omega_{~3}^{2} & =\sin(\tau+\phi)d\theta+\sin\theta\cos\theta\cos(\tau+\phi)(qd\tau-pd\phi)\ .
\end{aligned}
\end{align}
One can check 
\[
g^{\mu\nu}{e^{i}}_{\mu}{e^{j}}_{\nu}={\delta^{ij}},\qquad\delta_{ij}{e^{i}}_{\mu}{e^{j}}_{\nu}={g_{\mu\nu}}\ ,
\]

\[
\omega^{ij}=\varepsilon^{ijk}e_{k}-\tilde{\omega}^{ij} \ ,
\]
where 
\[
\tilde{\omega}^{ij}=\varepsilon^{ij3}\left(\left(q-1\right)d\tau+\left(p-1\right)d\phi\right)\ ,
\]
and 
\begin{align}
\nabla_{\mu}\varepsilon & =\left(\partial_{\mu}+\frac{1}{8}{\omega_{\mu}}^{ij}\left[\sigma_{i},\sigma_{j}\right]\right)\varepsilon=\left(\partial_{\mu}+\frac{i}{2}\gamma_{\mu}-\frac{i}{2}\left(\left(q-1\right){\delta_{\mu}}^{\tau}+\left(p-1\right){\delta_{\mu}}^{\phi}\right)\sigma_{3}\right)\varepsilon\ .
\end{align}
When $p=q=1$ we recover the round three-sphere. The usual left-invariant
basis is then defined by \eqref{eq:vielbein}.

\section{\label{sec:The-resolved-space}The resolved space}

For the resolved space \eqref{eq:smooth_two_parameter_metric} with
$f(\theta)\equiv f_{\epsilon}(\theta)$ we choose a vielbein 
\begin{align}
\begin{aligned}e^{1} & =\frac{1}{\sqrt{f(\theta)}}\sin(\tau+\phi)d\theta+\cos(\tau+\phi)\sin\theta\cos\theta(qd\tau-pd\phi)\ ,\\
e^{2} & =-\frac{1}{\sqrt{f(\theta)}}\cos(\tau+\phi)d\theta+\sin(\tau+\phi)\sin\theta\cos\theta(qd\tau-pd\phi)\ ,\\
e^{3} & =q\sin^{2}\theta d\tau+p\cos^{2}\theta d\phi\ ,
\end{aligned}
\label{eq:Resolved_space_vielbein}
\end{align}
with a spin connection 
\begin{align}
\begin{aligned}\omega_{~2}^{1} & =(1-q\sqrt{f(\theta)}\cos^{2}\theta)d\tau+(1-p\sqrt{f(\theta)}\sin^{2}\theta)d\phi\ ,\\
\omega_{~3}^{1} & =\cos(\tau+\phi)d\theta-\sqrt{f(\theta)}\sin\theta\cos\theta\sin(\tau+\phi)(qd\tau-pd\phi)\ ,\label{eq:Resolved_{s}pace_{s}pin_{c}onnection}\\
\omega_{~3}^{2} & =\sin(\tau+\phi)d\theta+\sqrt{f(\theta)}\sin\theta\cos\theta\cos(\tau+\phi)(qd\tau-pd\phi)\ .
\end{aligned}
\end{align}

The Ricci tensor of the resolved space is given by%
\footnote{Our convention for the Ricci scalar is that of \cite{Closset:2012ru}.
With this convention, $R$ on $S^{3}$ is negative.%
} 
\begin{align}
\begin{aligned}R_{\theta\theta} & =-2+\frac{\cot(2\theta)f'(\theta)}{f(\theta)}\ ,\\
R_{\tau\tau} & =-\frac{q^{2}}{2}\sin\theta(4\sin\theta f(\theta)-\cos\theta f'(\theta))\ ,\\
R_{\phi\phi} & =-\frac{p^{2}}{2}\cos\theta(4\cos\theta f(\theta)+\sin\theta f'(\theta))\ ,
\end{aligned}
\end{align}
and the Ricci scalar follows as 
\begin{align}
R=-6f(\theta)+2\cot(2\theta)f'(\theta)\ .
\end{align}
We can read off the form of the Ricci scalar in $\epsilon\to0$ limit
by integrating it on the resolved space 
\begin{align}
\begin{aligned}\int d^{3}x\sqrt{g}R & =4\pi^{2}pq\int_{0}^{\pi/2}d\theta\sin\theta\cos\theta\frac{1}{f^{1/2}(\theta)}(-6f(\theta)+2\cot(2\theta)f'(\theta))\ ,\\
 & =4\pi^{2}pq\Big[\int_{0}^{\epsilon}d\theta\left(6\theta f^{1/2}(\theta)-\frac{f'}{f^{1/2}}\right)+\int_{\epsilon}^{\pi/2-\epsilon}\sin\theta\cos\theta\cdot6\\
 & \qquad\qquad\qquad+\int_{\pi/2-\epsilon}^{\pi/2}\left(6\left(\frac{\pi}{2}-\theta\right)f^{1/2}(\theta)-\frac{f'}{f^{1/2}}\right)\Big]\ ,\\
 & =4\pi^{2}pq\left[3+2\left(\frac{1}{q}-1\right)-2\left(\frac{1}{p}-1\right)+O(\epsilon)\right]\ .
\end{aligned}
\end{align}
It follows that there are delta functional terms at $\theta=0$ and
$\theta=\pi/2$ in the Ricci scalar 
\begin{align}
R=-6-\frac{2}{\sin\theta\cos\theta}\left(\frac{1}{q}-1\right)\delta(\theta)+\frac{2}{\sin\theta\cos\theta}\left(\frac{1}{p}-1\right)\delta\left(\frac{\pi}{2}-\theta\right)\ .\label{eq:RSdelta}
\end{align}
Similarly, the Ricci tensor has delta functional terms 
\begin{align}
\begin{aligned}R_{\theta\theta} & =-2-\frac{1}{\sin\theta\cos\theta}\left(\frac{1}{q}-1\right)\delta(\theta)+\frac{1}{\sin\theta\cos\theta}\left(\frac{1}{p}-1\right)\delta\left(\frac{\pi}{2}-\theta\right)\ ,\\
R_{\tau\tau} & =-2q^{2}\sin^{2}\theta-q^{2}\tan\theta\left(\frac{1}{q}-1\right)\delta(\theta)\ ,\\
R_{\phi\phi} & =-2p^{2}\cos^{2}\theta+p^{2}\cot\theta\left(\frac{1}{p}-1\right)\delta\left(\frac{\pi}{2}-\theta\right)\ .
\end{aligned}
\label{RTdelta}
\end{align}

\section{\label{sec:Special-functions}Special functions}

The hyperbolic gamma function is a meromorphic function of a single
complex variable with two parameters defined in \cite{van2007hyperbolic}
\begin{align}
\begin{aligned}\Gamma_{h}\left(z;\omega_{1},\omega_{2}\right) & =\prod_{n_{1},n_{2}\ge0}\frac{(n_{1}+1)\omega_{1}+(n_{2}+1)\omega_{2}-z}{n_{1}\omega_{1}+n_{2}\omega_{2}+z}\ ,\\
 & =\exp\left(i\int_{0}^{\infty}\frac{dx}{x}\left(\frac{z-\omega}{\omega_{1}\omega_{2}x}-\frac{\sin(2x(z-\omega))}{2\sin(\omega_{1}x)\sin(\omega_{2}x)}\right)\right)\ ,
\end{aligned}
\end{align}
with the integral defined for 
\[
0<\Im\left(z\right)<\Im\left(\omega_{1}+\omega_{2}\right)\ ,
\]
and then analytically continued to the entire complex plain. The poles
are at 
\[
\Lambda=-\omega_{1}\mathbb{Z}_{\ge0}-\omega_{2}\mathbb{Z}_{\ge0}\ ,
\]
and the zeros at 
\[
\omega_{1}+\omega_{2}-\Lambda\ .
\]
We will sometimes suppress $\omega_{1,2}$ and denote $\Gamma(z)\equiv \Gamma(z;\omega_1,\omega_2)$. 
We also define 
\[
\omega=\frac{\omega_{1}+\omega_{2}}{2}\ .
\]

The function satisfies 
\begin{align}
\Gamma_{h}(z+\omega_{1}) & =2\sin\left(\frac{\pi z}{\omega_{2}}\right)\Gamma_{h}(z)\ ,\label{hgam}\\
\Gamma_{h}(z+\omega_{2}) & =2\sin\left(\frac{\pi z}{\omega_{1}}\right)\Gamma_{h}(z)\ ,\\
\Gamma_{h}(\omega_{1}+\omega_{2}-z) & =\Gamma_{h}(z)^{-1}\ ,\\
\Gamma_{h}(z;\omega_{1},\omega_{2}) & =\Gamma_{h}(\alpha z;\alpha\omega_{1},\alpha\omega_{2})\ ,\qquad\alpha\in\mathbb{C}\backslash\{0\}\ ,
\end{align}
and hence 
\[
\Gamma_{h}\left(\pm z\right):=\Gamma_{h}\left(z\right)\Gamma_{h}\left(-z\right)=\frac{-1}{4\sin(\pi z/\omega_{1})\sin(\pi z/\omega_{2})}\ .
\]
There are also multiple-angle formulas 
\begin{align}
\Gamma_{h}\left(Nz\right)=\prod_{k_{1},k_{2}=0,\cdots,N-1}\Gamma_{h}\left(z+\frac{k_{1}\omega_{1}+k_{2}\omega_{2}}{N}\right)\ .
\end{align}
For an integer $N\ge1$, it satisfies 
\begin{align}
\Gamma_{h}\left(z;i,\frac{i}{N}\right)=\prod_{k=0}^{N-1}\Gamma_{h}\left(z+\frac{ik}{N};i,i\right)\ .\label{HGidentity}
\end{align}
There is a product formula of dividing $N$ 
\begin{align}
\prod_{\substack{k_{1},k_{2}=0,\cdots,N-1\\
~(k_{1},k_{2})\neq(0,0)
}
}\Gamma_{h}\left(\frac{k_{1}\omega_{1}+k_{2}\omega_{2}}{N}\right)=\frac{1}{N}\ .
\end{align}
Some of the values at special points are 
\begin{align}
\begin{aligned}\Gamma_{h}\left(\omega\right)=1\ ,\qquad\Gamma_{h}\left(0\right) & =0\ ,\qquad\Gamma_{h}\left(\frac{\omega_{i}}{2}\right)=\frac{1}{\sqrt{2}}\ ,\qquad\Gamma_{h}\left(\omega+\frac{\omega_{i}}{2}\right)=\sqrt{2}\ ,\\
\Gamma_{h}\left(\omega_{1}\right) & =\sqrt{\frac{\omega_{1}}{\omega_{2}}}\ ,\qquad\Gamma_{h}\left(\omega_{2}\right)=\sqrt{\frac{\omega_{2}}{\omega_{1}}}\ .
\end{aligned}
\end{align}
$\Gamma_{h}$ is related to the double sine function $S_{2}$ and
to the double gamma function $\Gamma_{2}$ defined in \cite{kurokawa2003multiple}
by 
\[
\Gamma_{h}\left(z;\omega_{1},\omega_{2}\right)=S_{2}\left(z;\omega_{1},\omega_{2}\right)^{-1}=\frac{\Gamma_{2}\left(2\omega-z;\omega_{1},\omega_{2}\right)}{\Gamma_{2}\left(z;\omega_{1},\omega_{2}\right)}\ ,
\]
and to the Ruijsenaars' hyperbolic gamma function $G$ defined in \cite{ruijsenaars2000barnes}
by 
\[
\Gamma_{h}\left(z;\omega_{1},\omega_{2}\right)=G\left(-i\omega_{1},-i\omega_{2};z-\omega\right)\ .
\]

\bibliographystyle{JHEP}
\bibliography{Renyi}

\end{document}